\documentclass[12pt,draftclsnofoot,onecolumn]{IEEEtran}
\IEEEoverridecommandlockouts
\usepackage{cite}
\usepackage{amsmath,amssymb,amsfonts}
\usepackage{algorithmic}
\usepackage{graphicx}
\usepackage{textcomp}
\usepackage{tabularx}
\usepackage{scalerel}
\newcolumntype{L}{>{\raggedright\arraybackslash}X}
\usepackage{xcolor}
\usepackage{subfig}
\newtheorem{theorem}{Theorem}
\newtheorem{lemma}{Lemma}
\newtheorem{remark}{Remark}
\newtheorem{corollary}{Corollary}
\newtheorem{definition}{Definition}

\begin{document}
\title{Beam Management in 5G:\\ A Stochastic Geometry Analysis}

\author{Sanket S. Kalamkar, Fran\c{c}ois Baccelli, \\Fuad M. Abinader Jr.,  Andrea S. Marcano Fani, and Luis G. Uzeda Garcia
\thanks{F. Baccelli and S. S. Kalamkar are with INRIA-ENS, Paris, France. F. M. Abinader Jr., A. S.  Marcano Fani, and L. G. Uzeda Garcia are with Nokia Bell Labs, Paris, France. (e-mail: sanket.kalamkar.work@gmail.com, francois.baccelli@ens.fr, fuad.abinader@nokia-bell-labs.com, andrea.marcano@nokia-bell-labs.com, luis.uzeda\_garcia@nokia-bell-labs.com). This work was carried out at the Laboratory of Information, Networking and Communication Sciences (LINCS), Paris, France.}
\thanks{This work has received funding from the European Research Council (ERC)
under the European Union's Horizon 2020 research and innovation programme grant agreement number 788851.}
\thanks{Part of this work has been accepted in the 2020 IEEE Global Communications Conference (GLOBECOM'20)~\cite{bm_glo20}.}}

\maketitle

\begin{abstract}
Beam management is central in the operation of beamformed wireless cellular systems such as 5G New Radio (NR) networks. Focusing the energy radiated to mobile terminals (MTs) by increasing the number of beams per cell increases signal power and decreases interference, and has hence the potential to bring major improvements on area spectral efficiency (ASE). This paper proposes a first system-level stochastic geometry model encompassing major aspects of the beam management problem: frequencies, antenna configurations, and propagation; physical layer, wireless links, and coding; network geometry, interference, and resource sharing; sensing, signaling, and mobility management. This model leads to a simple analytical expression for the effective rate that the typical user gets in this context. This in turn allows one to find the number of beams per cell and per MT that maximizes the effective ASE by offering the best tradeoff between beamforming gains and beam management operational overheads and costs, for a wide variety of 5G network scenarios including millimeter wave (mmWave) and sub-6 GHz. As part of the system-level analysis, we define and analyze several underlying new and fundamental performance metrics that are of independent interest. The numerical results discuss the effects of different systemic tradeoffs and performance optimizations of mmWave and sub-6 GHz 5G deployments.
\end{abstract}


\section{Introduction}
The ever-increasing demand in capacity for mobile communications necessitates new implementation approaches that can significantly boost data rates and the area spectral efficiency (ASE) of mobile networks. One key enabler considered in 5G~\cite{5G_book} to face this demand is the use of the spectrum beyond the sub-6 GHz frequencies, known as millimeter wave (mmWave). More specifically, 5G is designed to make use of spectrum above 20 GHz, where bandwidth sizes of up to 400 MHz per carrier can be used to offer very high data rates (above 10 Gbps peak rates) and increase the network capacity~\cite{3gpp.38.101-2}; nevertheless, the sub-6 GHz bands, with up to 100 MHz of bandwidth per carrier, are still needed to ensure wide area coverage and data rates up to a few Gbps~\cite{3gpp.38.101-1}.

One of the key difficulties faced by 5G operation on mmWave frequencies is their challenging propagation characteristics: they are subject to high path loss, penetration loss, and diffraction loss, and are thus limited to short distance links, typically of a few hundred meters. But what was once considered a limitation makes nowadays mmWave a suitable candidate for small cells deployments, which can be used for network densification and capacity boosting.

To overcome the propagation challenges at mmWave frequencies, steerable antenna arrays with a large number of antenna elements each can be used to create highly directional beams that concentrate the transmitted energy to achieve high gains and increase coverage. Also, mmWave communications must be designed to operate under mobility conditions, covering users in LOS (Line-of-Sight) and NLOS (Non-Line-of-Sight) at pedestrian and vehicular speeds. This can become quite challenging since operation at mmWave frequencies is highly sensitive to changes in the environment (e.g., moving of the hand/head, passing cars). Thus, any mmWave-based wireless access system relies on techniques such as adaptive beamforming, beam-tracking and fast beam-switching to always maintain the best selected beam that maximizes the performance even in the face of beam misalignment events due to intra-cell mobility and inter-cell handovers. Although more important for mmWave frequencies, beamforming techniques with directional narrow beams can also be used in sub-6 GHz frequencies to improve the network performance.

\subsection{Motivation}
The introduction of beamforming as a central feature in 5G New Radio (NR) requires a series of beam management procedures to ensure efficient operation. One of such procedures is the so-called beam selection (or, beam refinement). Here the mobile terminal (MT) monitors beamformed reference signals (RSs) periodically transmitted by the base station (BS) to identify and report to the BS the \textit{best serving BS beam}. Such RSs can be either synchronization signal/physical broadcast channel blocks (SSBs), or channel state information reference signals (CSI-RSs). Another such procedure is the beam failure detection and recovery, where the MT monitors the radio link quality of the serving BS beam. If the quality of the radio link is repeatedly below a certain threshold, the BS triggers a procedure to quickly re-establish communication on neighboring beams.

Although beam management procedures are necessary for beamformed 5G NR operations, they also bring their own inherent costs in terms of signaling and delay overheads. For instance, there should be one beamformed BS RS for each beam in the beamset for a proper beam management operation. Hence, increasing the beamset size increases the signaling overhead. Thus the network planning for 5G NR deployments must take beam management into consideration. This requires the understanding of existing systemic effects and tradeoffs for any given MT mobility pattern, inter-site distance (also known as inter-BS distance), and the number/width of the beams. For example, the adoption of a BS beamset containing a large number of narrow beams during the network planning phase improves the signal-to-interference-plus-noise ratio (SINR) due to the increased signal power and reduced interference. But it may also lead to more frequent service interruption due to, e.g., beam switching delays (i.e., between the moment the BS RSs are measured and reported and the moment the BS decides to make the beam switch) and beam misalignment, which in the end may degrade the network performance. 

System-level simulations can be crucial to evaluate all such tradeoffs and determine the best combination of parameters for a given network deployment. But given the complex nature of beamforming techniques and beam management procedures in 5G NR, such simulations can be very time-consuming and expensive. This paper proposes a mathematical framework that permits a system-level analysis of beam management in 5G, leads to crisp insights into the associated tradeoffs, and has the potential to make beam management efficient. We believe that this mathematical framework will complement and help guide complex system-level simulations.

\subsection{Contributions}
This work makes the following contributions:
\begin{enumerate}
    \item We develop a first mathematical optimization framework for the beam management problem in 5G. Using the well-established tools from stochastic geometry~\cite{FB_book, MH_book,cell_book,rate_coverage}, our proposed optimization yields the number of beams at the BS and the MT that maximizes the effective ASE of downlink transmissions to the directional MT.
    \item The optimization permits a system-level analysis of 5G NR deployment based solely on a small set of input parameters. In particular, it takes into account the average speed of the MTs, mobility-induced beam misalignment, geometry-based beam selections during BS handovers and beam reselections within a cell, and time overheads associated with these beam (re)selections. For instance, given a mix of omnidirectional and directional MTs, a mix of MT speeds (pedestrians, bikes, and cars), a mix of geometries (some cells are bigger, other smaller, with the MTs either close or far from the serving BS), a mix of fading and blockages present in the network, our proposed optimization framework enables the identification of the number of beams that provides the best effective ASE, namely the best ``sum rate" for MTs of all types in a large domain.
    \item As part of the calculation of the effective ASE at the system level, we define and evaluate many new and fundamental performance metrics. These metrics are of independent interest and are given in Table~\ref{tab:2} with their equation numbers in the paper and their interpretation.
    \item While the proposed framework is generic and can cover a variety of related beam-centric systems such as IEEE 802.11ax/Wi-Fi 6, we give numerical results to evaluate a 5G NR-compliant radio access network (RAN) operating in a dense urban macro/picocell scenario, for both sub-6 GHz and mmWave frequencies. The results reveal the inter-dependencies between system parameters and provide insights into the associated tradeoffs.
    
\end{enumerate}

  \begin{table}

    \centering
   \caption{New Performance Metrics}
\begin{tabularx}{\linewidth}{|l|c|L|} 
    \hline
\textbf{Performance Metric} & \textbf{Formula} & \textbf{Interpretation of the Metric} \\ 
    \hline
Probability of beam misalignment              &  \eqref{eq:p_bm} and \eqref{eq:p_bm_MT}              & {The probability of beam misalignment between the beams of the BS and the MT due to the mobility of the MT between two consecutive beam measurement events.}                                        \\\hline
  Intensity of geometry-based beam reselection             & \eqref{eq:beam_resel_int} &  Frequency of geometry-based beam switching within the cells. \\\hline
    Geometry-based time-of-stay           & \eqref{eq:time_stay} &  Average time spent by the MT within the main lobe of a beam. \\\hline
  
Effective intensity of  beam reselection             & \eqref{eq:eff_int} &  Frequency of beam reselections within a cell by taking into account the geometry and the system design. It determines the time overhead associated with beam reselection.\\\hline
Effective time-of-stay   &  \eqref{eq:eff_tos}               &  Average duration of time intervals during which the MT has selected any given serving BS beam, which is essential to, e.g., evaluate whether the MT will have enough time to perform MT beam refinement and link adaptation before data transmission.  \\\hline

Probability of a beam switch          &  \eqref{eq:prob_sw} &           At a cell boundary, it is the probability that the MT has to switch panels/beams. This allows one to predict the \textit{no} beam switching event, which can help reduce the number of beam measurements and the associated time overhead.                      \\\hline

 Time overheads  &  \eqref{eq:no_o}, \eqref{eq:full_o}, \eqref{eq:partial_o}                 &  The time overheads are associated with the MT mobility and reference signal measurements in 5G that are necessary for inter-cell handovers and beam switching within the cell. The overheads determine the effective time available for data transmission.                                        \\\hline
\end{tabularx}
\label{tab:2}
    \end{table}

\subsection{Related Work}
Stochastic geometry has been widely used to model and analyze several aspects of wireless networks including cellular networks~\cite{FB_book, MH_book, cell_book, rate_coverage}. A stochastic geometry framework is also popular to model mmWave cellular networks and evaluate key performance aspects such as initial access, success probability, the average rate of the typical user, and the effect of blockages~\cite{Bai, Renzo, Li_IA, Jain}.

The study of handovers due to user mobility in non-beam-centric cellular networks and their effect on various performance metrics such as throughput is a rich area of research~\cite{Survey_handover1, Survey_handover2, Tabassum}. Most of the works have focused on BS handovers (or, equivalently, cell handovers)~\cite{Tokuyama,Bartek,Teng,Arshad,Arshad2,Demarchou}, and they ignore the effect of these handovers on beam-based communications. The work in~\cite{Li} is probably the closest one to our work. This work studies both initial beam selections during BS handovers and beam reselections within a cell and their associated overheads in a beam-centric mmWave cellular network. The work in \cite{Li} obtains analytical expressions of inter-cell and inter-beam handover rates based on which the ASE is calculated subject to overheads due to these handovers. But, it considers only noise-limited scenarios and ignores interference from other BSs, while our work considers both noise and interference in the analysis. Second, our work also studies the effect of handovers on optimizing the number of beams during network planning phase, while \cite{Li} assumes a fixed number of beams. Third, unlike \cite{Li}, we consider blockages and beam misalignment due to mobility. Finally, we also consider the directional reception at MTs and its effect on beam management. Overall, there is a lack of understanding on how network planning decisions (such as BS beamset and inter-site distance), beam management procedures (with its associated overhead) and MT mobility affects system-level performance in the presence of interference, fading, blockages, and beam misalignment error. The present paper can be seen as an attempt to incorporate all this above in a system-level model.

\begin{table}[!t]
\centering
\caption{Notation}
\begin{tabularx}{\linewidth}{|l|L|l|L|} 
\hline
  \textbf{Symbol}              &  \textbf{Definition}  & \textbf{Symbol} & \textbf{Definition}                                                              \\\hline
$\lambda$   &  BS intensity & $v$ & MT speed                                                     \\\hline
$f_{\rm c}$ & Carrier frequency & $c$ & Speed of light \\\hline
$2^n$            &  Number of BS beams & $2^k$ & Number of MT beams                                                      \\\hline
$\varphi_n$ & Beamwidth at the BS & $\varphi_k$ & Beamwidth at the MT 
\\\hline
$G_{{\rm m}, n}$, $G_{{\rm s}, n}$ & Main and side lobe gains at the BS & $G_{{\rm m}, k}$, $G_{{\rm s}, k}$ & Main and side lobe gains at the MT\\\hline
$p_{\rm bm}^{\rm BS}$ & Beam misalignment probability at the BS & $p_{\rm bm}^{\rm MT}$ & Beam misalignment probability at the MT\\\hline
$\tau$ & SSB periodicity & $R_{\rm c}$ & LOS ball radius\\\hline
$\alpha_{\rm L}$ & LOS path-loss exponent & $\alpha_{\rm N}$ & NLOS path-loss exponent \\\hline
$W$ & Bandwidth & $N_0$ & Noise power density \\\hline
$P$ & Transmit power & $\sigma^2$ & Noise power \\\hline
$\beta$ & SINR threshold & $Q_{\rm max}$ & Maximum SINR \\\hline
$\mathcal{R}$ & Ergodic Shannon rate & $\mathcal{R}_{\rm eff}$ & Effective ASE \\\hline
$T_{\rm c}$ & Cell handover overhead & $T_{\rm {b}}$ & Beam reselection overhead \\\hline
$\mu_{\rm{s, c}}$ & Linear intensity of BS handovers & $\mu_{\rm c}$ & Time intensity of BS
handovers \\\hline
$\mu_{\rm{s,b}}$ & Linear intensity of geometry-based beam reselections & $\mu_{\rm t, b}$ & Time intensity of geometry-based beam reselections \\\hline
$\mu_{\rm b}$ & Effective intensity of beam reselections & $T_{\rm o}$ & Average time overhead per unit time \\\hline 
$T_{\rm st}$ & Geometry-based time-of-stay & $T_{\rm eff, st}$ & Effective time-of-stay \\\hline
$p_{\rm sw}$ & Probability of a beam switch &  $p_{\rm s}$ & Success probability of a transmission\\\hline

\end{tabularx}
\label{tab:notation}
\end{table}

\subsection{Organization of the Paper}
 Section~\ref{sec:sys_model} gives the models for network geometry, beamforming, channel, and blockages and introduces the mobility-induced beam misalignment. Section~\ref{sec:rate} calculates simple analytical expressions for the ergodic Shannon rate for sub-6 GHz and mmWave deployments. Section~\ref{sec:beam_reselection} discusses the beam selection during an inter-cell handover and calculates the intensity of beam reselections within the cell. This section also calculates the time overheads associated with these beam (re)selections. Section~\ref{sec:effective_ASE} gives the final optimization problem of maximizing the effective ASE. Section~\ref{sec:network_setup} provides the network setup with the values of network parameters. Sections~\ref{sec:omni_result} and \ref{sec:direct_result} discuss several systemic tradeoffs for omnidirectional and directional MTs. Section~\ref{sec:new_metrics} discusses briefly some new performance metrics mentioned in Table~\ref{tab:2}. Section~\ref{sec:conclusions} gives conclusions and future directions. Table~\ref{tab:notation} provides the notation used in the paper.
 \begin{figure}
\centering
\includegraphics[scale=.68]{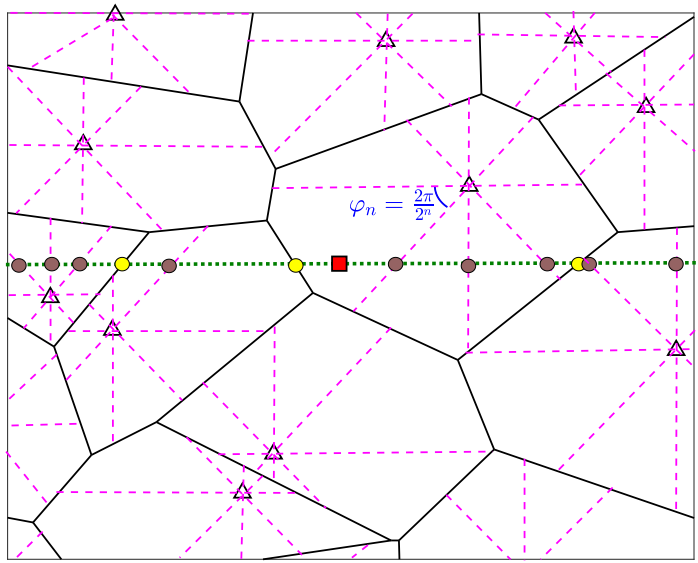}
\caption{A snapshot of a Poisson cellular network with directional beamforming to an MT. Here, each BS has $2^3= 8$ beams, i.e., $n = 3$. Triangle : Base station (BS), red square : MT location, brown circle : Geometry-based beam reselection location, yellow circle : BS handover location, solid black lines : Cell boundaries, dashed lines : Beam boundaries, and dotted line : Trajectory of the MT.}
\label{fig:PVC}
\end{figure}

\section{System Model}
\label{sec:sys_model}
\subsection{Network Model}
As shown in Fig.~\ref{fig:PVC}, we consider a downlink cellular network, where the BS locations are modeled as a homogeneous Poisson point process (PPP) $\Phi \subset \mathbb{R}^2$ with intensity $\lambda$. We assume that a directional MT moves on a randomly oriented straight line with speed $v$. Without loss of generality, thanks to the isotropy and the stationarity of the PPP~\cite{FB_book}, this line of MT motion can be considered to be along the $x$-axis and passing through the origin. We assume that each BS always has an MT to serve. Also, a BS serves one MT at a time per resource block.\footnote{The analysis in this paper can be extended to the case where a resource block at a BS is shared by multiple MTs.} The MT associates itself with the nearest BS. Such an association results in BS cells forming a Poisson-Voronoi tessellation as shown in Fig.~\ref{fig:PVC}. Let $X_0 \in \Phi$ denote the location of the serving BS of the MT at a given time. Without loss of generality, we can focus on the MT that is located at the origin at that time. After averaging over the PPP, this MT becomes the \textit{typical} MT.

\subsection{Beamforming Model}

A BS at location $X \in \Phi$ uses directional beamforming to communicate with the typical directional MT located at the origin. As shown in Fig.~\ref{fig:PVC}, we approximate the actual antenna pattern by a sectorized one, where each sector corresponds to the main lobe of one beam of the BS. For simplicity, we assume that each BS has $2^n$ beams with $n \in \mathbb{N}$, which corresponds to $2^{n-1}$ beam boundaries.\footnote{Here, a beam boundary is a line segment that connects two points on the cell boundary and passes through the location of the BS, as shown in Fig.~\ref{fig:PVC}.} Hence, the beamwidth for a BS is
\begin{align}
\varphi_n &= \frac{2\pi}{2^n} = \frac{\pi}{2^{n-1}}.
\label{eq:bw}
\end{align} 
We focus on a simple antenna pattern model where the main lobe is restricted to the beamwidth. Both the main and side lobe gains depend on the number $2^n$ of beams. In particular, the antenna gain $G_n$ is expressed as
\begin{align*}
G_n(\psi) = \begin{cases}
               G_{{\rm m},  n}  & \text{if}~|\psi| \leq \varphi_n/2\\
               G_{{\rm s},  n} & \text{otherwise},
            \end{cases}
\end{align*}
where $G_{{\rm m},  n}$ and $G_{{\rm s},  n}$ denote the antenna gains within the main lobe and the side lobe, respectively, as a function of the number $2^n$ of beams, and $\psi$ is the angle off the boresight direction.

The beamforming model at the MT is the same as the one at the BS, where the number of beams at the typical MT is $2^k$ with $k \in \mathbb{N}$. Thus the antenna gain $G_k$ at the typical MT is expressed as 
\begin{align*}
G_k(\psi) = \begin{cases}
               G_{{\rm m},  k}  & \text{if}~|\psi| \leq \varphi_k/2\\
               G_{{\rm s},  k} & \text{otherwise},
            \end{cases}
\end{align*}
where $\varphi_k = \frac{\pi}{2^{k-1}}$ is the beamwidth at the typical MT.

\subsection{Beam (Re)Selection}
\label{sec:beam_resel_intro}
As shown in Fig.~\ref{fig:PVC}, the MT may cross several cell and beam boundaries as it moves on a straight line. At a cell boundary, a BS handover occurs. Thus, a new beam at the next serving BS and potentially at the MT need to be selected during a BS handover.

\begin{figure}
\centering
\includegraphics[scale=.42]{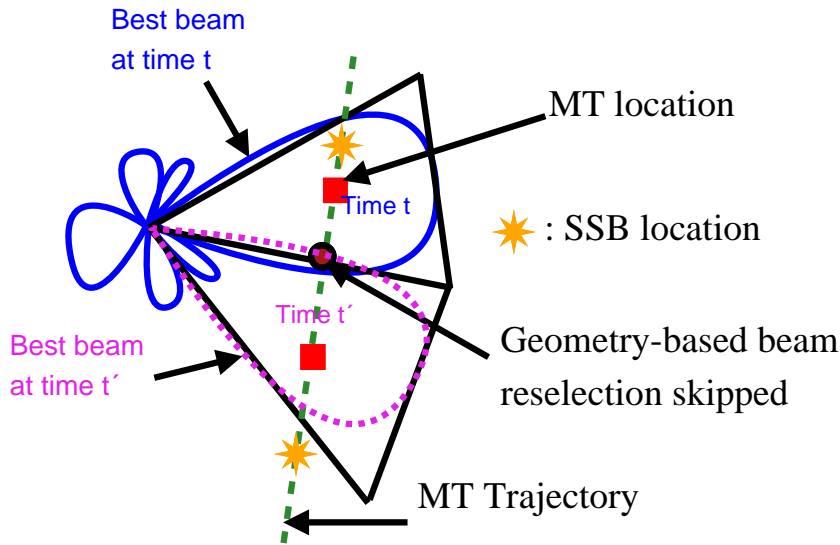}
\caption{Mobility-induced beam misalignment. Ideally, the beam reselection should happen at the beam boundary for the MT to stay connected to the main lobe of the \textit{best} beam. But, since the SSB (a beam-formed reference signal) location might not coincide with the location of ideal beam reselection at the beam boundary, the MT at time $t'$ is still connected to the side lobe of the previously selected beam at time $t$, which results in a beam misalignment.}

\label{fig:tos}
\end{figure}

Within the cell as well, the MT moves from the main lobe of one beam to that of another beam. Thus, a new beam has to be \textit{reselected} at the BS and potentially at the MT as well in order for the MT to stay connected with the \textit{best serving BS} beam.\footnote{\label{foot:best_beam}The best beam is the beam of the serving BS that has the MT in its main lobe. For example, as shown in Fig.~\ref{fig:tos}, at time $t$, the beam represented by the solid blue line is the best beam, while at time $t'$, the beam represented by the dashed pink line is the best beam. Hence, the best beam selection is purely \textit{geometry-based} and ignores reflections of signals from the BS in order to maintain the focus of the paper on the most fundamental aspects of beam management.} Ideally, the herein called \textit{geometry-based beam reselection} would take place, where in order for the MT to stay connected with the main lobe of the best beam (as defined in Footnote~\ref{foot:best_beam}), it should reselect the beam at a beam boundary. Such a location of beam reselection is marked by ``brown circle" in Fig.~\ref{fig:tos}. However, due to 5G NR system characteristics the herein called \textit{measurement-based beam reselection} takes place, where beam reselection events only occur after beam measurement events of SSB bursts~\cite{tut_bm}, which are transmitted periodically by the BS with period $\tau$.

Hence the overall beam reselection process considers both the geometry- and measurement-based beam reselection events. However, the mismatch between the locations of geometry- and measurement-based beam reselections may lead to the beam misalignment, which we discuss in the following subsection using Fig.~\ref{fig:tos}.

\subsection{Mobility-induced Beam Misalignment}
Beam misalignment occurs when the main lobes of the serving BS and the typical MT are not aligned with each other. Specifically, during an SSB, at the serving BS, the beam that has the MT within its main lobe (i.e., the best beam) is selected for communication. We call this beam the \textit{reference beam} (see Fig.~\ref{fig:tos}). Due to the mobility of the MT, if the duration between two consecutive SSBs is sufficiently long, it is possible that the MT has moved out of the main lobe of the reference beam without selecting the new beam at the beam boundary (i.e., without performing geometry-based beam reselection).\footnote{The duration between two consecutive SSBs is a system design parameter. It this duration is very small, we get the geometry-based beam reselection.} Thus, the MT is still connected to the reference beam via side lobe until the next SSB when the new reference beam is selected.  Such a mobility-induced beam misalignment at the BS reduces the signal strength at the MT as it then lies within the side lobe of the reference beam selected during the previous SSB.

Similar to the beam misalignment at the BS, there can be beam misalignment at the typical MT. During an SSB, the beam that is pointing towards the serving BS is selected at the MT to receive from the serving BS. Due to the mobility of the MT between two consecutive SSBs, it is possible that the beam selected at the MT during the previous SSB is no more pointing its main lobe towards the serving BS. As a result, the MT receives from the serving BS in the side lobe of the beam that was selected earlier in the previous SSB. In the worst-case of beam misalignment, the side lobe of the selected beam at the serving BS is aligned with the side lobe of the selected beam at the typical MT.

At a given time, the beam misalignment probability $p_{\rm bm}^{\rm{BS}}$ at the BS depends on the speed $v$ of the MT, the duration $\tau$ between two SSBs, and the average distance between two  consecutive BS's beam boundaries. Here, we propose a simple yet effective formula that captures the effects of these parameters on the probability of beam misalignment at the BS due to MT mobility in a snapshot of the network. We evaluate the probability $p_{\rm bm}^{\rm{BS}}$ that the MT is outside the main lobe of the reference beam by the:
\begin{align}
p_{\rm bm}^{\rm{BS}} = 1 - \exp\left(-\frac{v\tau}{1/\mu_{\rm s,b}}\right),
\label{eq:p_bm}
\end{align}
where $1/\mu_{\rm s,b} = \frac{\pi}{2^n \sqrt{\lambda}}$ is the average distance between two consecutive BS's beam boundaries, with $\mu_{\rm s,b}$ being the linear intensity of beam boundary crossings given in Theorem~\ref{thm:int_beam} in Section~\ref{sec:beam_resel}.
We can interpret \eqref{eq:p_bm} as
\begin{align*}
1- p_{\rm bm}^{\rm{BS}} = \mathbb{P}(T > \tau),
\end{align*}
where $T$ is an exponential random variable with mean $\frac{1}{v\mu_{\rm s,b}}$, with $T$ interpreted as the time between two consecutive SSBs. Note that $\frac{1}{v\mu_{\rm s,b}} $ is the average time needed by the typical MT to travel between two consecutive BS's beam boundaries. Hence, the average time  between two SSBs is equal to the average time taken by the typical MT to travel between two consecutive BS's beam boundaries.

As in the case of beam misalignment on BS's side, we use a similar formula to calculate the probability $p_{\rm bm}^{\rm MT}$ of beam misalignment on MT's side as
\begin{align}
p_{\rm bm}^{\rm MT} = 1 - \exp\left(-\frac{v\tau}{1/\mu_{\rm s,m}}\right) =  1 - \exp\left(-\frac{2^k\sqrt{\lambda} v\tau}{\pi}\right),
\label{eq:p_bm_MT}
\end{align} 
where $1/\mu_{\rm s,m} = \frac{\pi}{2^k \sqrt{\lambda}}$ is the average distance between two consecutive MT's beam boundaries. 

Based on the beam misalignment at both the serving BS and the typical MT, the antenna gain $G_{0}$ of the serving BS is given by
\begin{align}
G_{0}(n, k) = \begin{cases}
               G_{{\rm m}, n} G_{{\rm m}, k} & \mathrm{w.p.}~\left(1-p_{\rm bm}^{{\rm BS}}\right)\left(1-p_{\rm bm}^{{\rm MT}}\right)\\
               G_{{\rm m}, n} G_{{\rm s}, k}  & \mathrm{w.p.}~\left(1-p_{\rm bm}^{{\rm BS}}\right)p_{\rm bm}^{{\rm MT}} \\
               G_{{\rm s}, n} G_{{\rm m}, k} & \mathrm{w.p.}~p_{\rm bm}^{{\rm BS}}\left(1-p_{\rm bm}^{{\rm MT}}\right)\\
               G_{{\rm s}, n} G_{{\rm s}, k}  & \mathrm{w.p.}~p_{\rm bm}^{{\rm BS}}p_{\rm bm}^{{\rm MT}}.
            \end{cases}
            \label{eq:gain_serv}
\end{align}

\subsection{Blockage and Channel Model}

In mmWave bands, the presence of blockages leads to LOS and NLOS propagation between the MT and a BS. We adopt a LOS ball model~\cite{Bai} to capture the effect of blockages. The propagation between a BS and the typical MT separated by distance $d$ is LOS if $d < R_{\rm c}$
where $R_{\rm c}$ is the maximum distance for LOS propagation. 
The LOS and NLOS channel conditions induced by the blockage effect are characterized by different path-loss exponents, denoted by $\alpha_{\rm L}$ and $\alpha_{\rm N}$, respectively. As shown in~\cite{Rappaport}, typical values of these path-loss exponents are $\alpha_{\rm L} \in [1.8, 2.5]$ and $\alpha_{\rm N} \in [2.5, 4.7]$.

The channel follows Rayleigh fading with unit mean power gain.\footnote{Although the LOS mmWave channels are better modeled by Nakagami-$m$ fading, Rayleigh fading allows one much better analytical tractability.  Also, simulation results in Section~\ref{sec:mix_MT} (see Figs.~\ref{fig:Ray_Naka_ISD_75_vel_30} and \ref{fig:Ray_Naka_ISD_250_vel_30}) show that Rayleigh and Nakagami fading yield the same trends in terms of effective ASE.} Let $h_{X}$ denote the channel power gain from the BS at location $X \in \Phi$ to the MT. Note that, the random variables $h_{X}$ are i.i.d. exponential random variables with unit mean, i.e., $h_{X} \sim \exp(1)$. Let $|X|$ denote the distance between a BS at $X \in \Phi$ and the typical MT located at the origin. We consider the standard power-law path-loss model with path-loss function
\begin{align}
l(X) = \begin{cases}
               K|X|^{-\alpha_{\rm L}} & \mathrm{if}~|X| < R_{\rm c}\\
               K|X|^{-\alpha_{\rm N}} & \mathrm{if}~|X| \geq R_{\rm c},
            \end{cases}
\end{align}
where $K = \left(\frac{c}{4\pi f_{\rm c}}\right)^2$ with $c$ the speed of light and $f_{\rm c}$ the carrier frequency.

On the other hand, the signals in sub-6 GHz bands experience more scattering and diffraction against to blocking and reflection in mmWave bands. Thus the effect of blockages is insignificant. Hence, when analyzing performance in sub-6 GHz bands, one may neglect blockages. This is equivalent to a special case of mmWave bands where one assumes only NLOS propagation between the typical MT and all BSs with the path-loss exponent between 2.5 and 4.7.

\subsection{Signal-to-Interference-Noise Ratio (SINR)}

When the typical MT is associated with the BS located at $X_0 \in \Phi$ with antenna gain $G_0$ as in \eqref{eq:gain_serv}, the SINR at the typical MT is given by
\begin{align}
\mathsf{SINR}_{n,k} = \frac{PG_{0}(n,k)h_{X_0} l(|X_0|)}{\sigma^2 + I_{n,k} }, 
\label{eq:sir}
\end{align}
where $P$ is the transmit power of the BS.  Also, $\sigma^2 = WN_0$ is the noise power, where $W$ and $N_0$ are the bandwidth and the noise spectral density, respectively. In the denominator of \eqref{eq:sir},  $I_{n,k}$ is the interference power at the typical MT given by\footnote{The framework can also be extended to the scenario where there is a frequency reuse among BSs as described in~\cite{rate_coverage}.}
\begin{align}
I_{n,k} = \sum_{X \in \Phi\setminus \lbrace X_0\rbrace} P G_X(n,k)h_{X}l(|X|),
\label{eq:intf}
\end{align}
where $G_X(n,k)$ is the antenna gain of an interfering BS located at $X \in \Phi$.

Since the beams of all BSs are oriented towards their respective MTs, the direction of arrivals between interfering BSs and the typical MT is distributed uniformly in $[-\pi, \pi]$. Thus, the gain $G_X$ of an interfering BS at $X \in \Phi$ is an i.i.d. discrete random variable with the probability mass function as:
\begin{align}
G_{X}(n, k) = \begin{cases}
               G_{{\rm m}, n} G_{{\rm m}, k} & \mathrm{w.p.}~p_{\rm mm} = \frac{\varphi_n \varphi_k}{4\pi^2} \\
               G_{{\rm m}, n} G_{{\rm s}, k}  & \mathrm{w.p.}~p_{\rm ms} =  \frac{\varphi_n (2\pi -\varphi_k)}{4\pi^2}\\
               G_{{\rm s}, n} G_{{\rm m}, k} & \mathrm{w.p.}~p_{\rm sm} = \frac{(2\pi -\varphi_n) \varphi_k}{4\pi^2} \\
               G_{{\rm s}, n} G_{{\rm s}, k}  & \mathrm{w.p.}~p_{\rm ss} = \frac{(2\pi -\varphi_n) (2\pi -\varphi_k)}{4\pi^2},
            \end{cases}
            \label{eq:gain_intf}
\end{align}
where $\varphi_n = \frac{\pi}{2^{n-1}}$ and $\varphi_k = \frac{\pi}{2^{k-1}}$ are the beamwidths of the BS and the typical MT, respectively.

\section{Ergodic Shannon Rate}
\label{sec:rate}
The downlink ergodic Shannon rate at the typical MT is given as
\begin{align*}
\mathcal{R}(n, k) = W\mathbb{E}\left[\log\left(1 + \mathsf{SINR}_{n,k}\right)\right],
\end{align*}
where $\mathbb{E}(\cdot)$ denotes expectation.

\begin{definition}[Success Probability]
The success probability $p_{\rm s}$ of the typical MT is the probability that the SINR at the typical MT exceeds a predefined threshold. Mathematically,
\begin{align*}
p_{\rm s}(n,k, \beta) \triangleq \mathbb{P}(\mathsf{SINR}_{n,k} > \beta),
\end{align*}
where $\beta > 0$ is the predefined SINR threshold, which also parametrizes the transmission rate.
\end{definition}

\begin{lemma}
\label{lem:suc_prob}
Let
\begin{align}
&F(\alpha_{\rm S}, \alpha_{\rm I}, w, G_0) \nonumber \\
&\triangleq 2\pi \lambda \left(1- \frac{p_{\rm mm}}{1+\frac{\beta r^{\alpha_{\rm S}} G_{{\rm m}, n} G_{{\rm m}, k} }{G_0 w^{\alpha_{\rm I}}}} -  \frac{p_{\rm ms}}{1+\frac{\beta r^{\alpha_{\rm S}} G_{{\rm m}, n} G_{{\rm s}, k} }{G_0 w^{\alpha_{\rm I}}}} -  \frac{p_{\rm sm}}{1+\frac{\beta r^{\alpha_{\rm S}} G_{{\rm s}, n} G_{{\rm m}, k} }{G_0 w^{\alpha_{\rm I}}}} -  \frac{p_{\rm ss}}{1+\frac{\beta r^{\alpha_{\rm S}} G_{{\rm s}, n} G_{{\rm s}, k} }{G_0 w^{\alpha_{\rm I}}}} \hspace{-1mm}\right)w.
\label{eq:F}
\end{align}
The success probability $p_{\rm s}(n,k,\beta)$ is 
\begin{align}
p_{\rm s}(n,k,\beta) &= \left(1-p_{\rm bm}^{\rm BS}\right)\left(1-p_{\rm bm}^{\rm MT}\right) q_{\rm s}(n,k,\beta, G_{{\rm m},n}G_{{\rm m},k})+\left(1-p_{\rm bm}^{\rm BS}\right)p_{\rm bm}^{\rm MT}q_{\rm s}(n,k,\beta, G_{{\rm m},n}G_{{\rm s},k}) \nonumber \\
&+ p_{\rm bm}^{\rm BS}\left(1-p_{\rm bm}^{\rm MT}\right) q_{\rm s}(n,k,\beta, G_{{\rm s},n}G_{{\rm m},k})+ p_{\rm bm}^{\rm BS}p_{\rm bm}^{\rm MT}q_{\rm s}(n,k,\beta, G_{{\rm s},n}G_{{\rm s},k}),
\label{eq:suc_prob_main}
\end{align}
where
\begin{align}
&q_{\rm s}(n,k,\beta, G_0) \nonumber \\
&= \int_{0}^{R_{\rm c}}f_{R}(r) \exp\left(-\frac{\beta r^{\alpha_{\rm L}}\sigma^2}{PKG_0}\right) \exp\left(\hspace*{-1mm}-\int_{r}^{R_{\rm c}}\hspace*{-1mm}F(\alpha_{\rm L}, \alpha_{\rm L}, w, G_0) \mathrm{d}w -\hspace*{-1mm} \int_{R_{\rm c}}^{\infty}\hspace*{-1mm}F(\alpha_{\rm L}, \alpha_{\rm N}, w, G_0)\mathrm{d}w\right) \mathrm{d}r \nonumber \\
&+ \int_{R_{\rm c}}^{\infty}\exp\left(-\frac{\beta r^{\alpha_{\rm N}}\sigma^2}{PKG_0} - \int_{r}^{\infty} F(\alpha_{\rm N}, \alpha_{\rm N}, w, G_0)\mathrm{d}w\right)f_{R}(r)\mathrm{d}r
\label{eq:suc_prob}
\end{align}
with $f_R(r) = 2\pi \lambda r e^{-\lambda \pi r^2}$ where $R$ is the distance to the nearest BS.
\end{lemma}
\begin{IEEEproof}
The proof is given in Appendix~\ref{app:suc_prob}.
\end{IEEEproof}

\begin{lemma}
For sub-6 GHz band, the success probability $p_{\rm s}$ is given by \eqref{eq:suc_prob_main} with
\begin{align}
    q_{\rm s}(n, k, \beta, G_0) =  \int_{0}^{\infty}f_{R}(r) \exp\left(-\frac{\beta r^{\alpha_{\rm L}}\sigma^2}{PKG_0}\right) \exp\left(-\int_{r}^{\infty}F(\alpha_{\rm N}, \alpha_{\rm N}, w, G_0) \mathrm{d}w \right) \mathrm{d}r. 
\end{align}
\end{lemma}
\begin{IEEEproof}
The desired expression of the success probability is obtained by substituting $R_{\rm c} = \infty$ and replacing $\alpha_{\rm L}$ with $\alpha_{\rm N}$ in \eqref{eq:suc_prob}.
\end{IEEEproof}

\begin{theorem}
Imposing the limit $Q_{\rm max}$ on the maximum achievable SINR stemming from RF imperfections and modulation schemes, the ergodic Shannon rate per unit time is
\begin{align}
\mathcal{R}(n, k)  = W\int_{0}^{Q_{\rm max}} \frac{p_{\rm s}(n,k,z)}{z+1}~\mathrm{d}z,
\label{eq:av_rate}
\end{align}
where $p_{\rm s}(n,k,z)$ is given by \eqref{eq:suc_prob}.
\end{theorem}
\begin{IEEEproof}
The proof follows directly from
\begin{align}
\mathcal{R}(n, k) &= W \int_{0}^{\infty} \mathbb{P}(\log(1+\min(\mathsf{SINR}_{n,k}, Q_{\rm max})) > z)~\mathrm{d}z \nonumber \\
&= W\int_{0}^{Q_{\rm max}} p_{\rm s }(n,k,e^z-1)~\mathrm{d}z \nonumber \\
&= W\int_{0}^{Q_{\rm max}} \frac{p_{\rm s}(n,k, z)}{z+1}~\mathrm{d}z.
\end{align}
\end{IEEEproof}

\section{Beam (Re)Selection}
\label{sec:beam_reselection}

\subsection{Beam Selection during BS Handovers}
When the MT performs a BS handover, a beam is selected with the new BS as well as the typical MT. We know from \cite{FB_cell_crossings} that, for the Poisson-Voronoi tessellation, the linear intensity of \textit{cell} boundary crossings, i.e., BS handovers, is $\mu_{\rm s,c} = \frac{4\sqrt{\lambda}}{\pi}$. Hence, the time intensity of BS handover (or, equivalently, beam selection) is
\begin{align}
\mu_{\rm c} = \frac{4\sqrt{\lambda}}{\pi}v.
\end{align}

\subsection{Beam Reselection Within the Cell}
\label{sec:beam_resel}
We now calculate the average number of beam reselections the typical MT performs per unit length and the time intensity of beam reselection, i.e., the average rate of beam reselections. As discussed in Section~\ref{sec:beam_resel_intro}, the beam reselection events are a combination of geometry- and measurement-based beam reselection events.

Recall that a geometry-based beam reselection occurs at a beam boundary within the Voronoi cell of a BS. In Fig.~\ref{fig:PVC}, the locations of geometry-based beam reselections are denoted by brown circles.

\begin{theorem}[Intensity of geometry-based beam reselection] 
For $2^n$ beams at a BS and PPP of intensity $\lambda$, the linear intensity $\mu_{\rm s,b}$ of geometry-based beam reselection for the typical MT moving on a straight line with speed $v$ is $\frac{2^n\sqrt{\lambda}}{\pi}$, while the time intensity $\mu_{\rm t,b}$ is 
\begin{align}
\mu_{\rm t,b}(n) = \frac{2^n\sqrt{\lambda}}{\pi}v.
\label{eq:beam_resel_int}
\end{align}
\label{thm:int_beam}
\end{theorem}\vspace{-6mm}
\begin{IEEEproof}
The proof is given in Appendix~{\ref{app:int_beam}}. 
\end{IEEEproof}

This theorem is valid when integrating over all motion directions of the MT chosen uniformly at random, and it does not hold for a fixed direction of MT motion.

\begin{corollary}
When taken into account the average number of geometry-based beam reselections that are skipped between two consecutive SSBs, the effective time intensity of beam reselection becomes
\begin{align}
\mu_{\rm b}(n) =  \frac{1}{\max\left(\tau, \frac{1}{\mu_{\rm t, b}(n)}\right)},
\label{eq:eff_int}
\end{align} 
where $\tau$ is the SSB periodicity.
\end{corollary}

\begin{corollary}
The geometry-based time-of-stay $T_{\rm st}$ of the MT defined as the average time spent by the MT within the main lobe of a beam is simply the inverse of the time intensity $\mu_{\rm t, b}$, i.e.,
\begin{align}
T_{\rm st} = \frac{1}{\mu_{\rm t, b}} = \frac{\pi }{2^n\sqrt{\lambda} v}.
\label{eq:time_stay}
\end{align}
\end{corollary}

\begin{corollary} The effective time-of-stay $T_{\rm eff, st}$ defined as the average time for which the MT is connected with a beam of the BS is 
\begin{align}
T_{\rm eff, st} = \frac{1}{\mu_{\rm b}(n) } = \max\left(\tau, \frac{1}{\mu_{\rm t, b}(n)}\right).
\label{eq:eff_tos}
\end{align}
\end{corollary}

\begin{remark}
The effective time-of-stay allows one to evaluate whether the MT will have enough time to perform MT beam refinement and link adaptation before data transmission. Also, for a small SSB periodicity $\tau$, the effective time-of-stay is equal to the geometry-based time-of-stay.
\end{remark}

\subsection{Time Overheads due to Beam (Re)Selection}
\label{sec:cases_overhead}
This subsection builds on the previous subsections on analytical results for beam (re)selections and incorporates the practical time overheads that arise from the 5G NR design.

BS handovers and beam reselections may result in significant overheads in terms of additional delay required for periodically conducting the SSB measurement and report which are required for the SSB beam sweeping and beam alignment procedures. Such an overhead in time reduces the overall time available for data transmissions, in turn reducing the ergodic Shannon rate.

The two components that contribute to the time overhead are:

\begin{enumerate}
\item The time $T_{\mathrm{c}}$ for executing the simultaneous cell and beam reselection procedures during each BS handover at \textit{cell} boundary crossing. This includes the periodic SSB measurement, receiver processing time for the SSBs, and the handover interruption time due to cell switch and radio resource control (RRC) reconfiguration.
\item The time $T_{\mathrm{b}}$ for beam alignment after each beam reselection within the cell of a BS, which includes the periodic SSB measurement and the receiver processing time for the SSBs.

\end{enumerate}

There is also a scaling factor for those SSB measurement delays, associated with the way the MT implements the SSB measurements, which is intrinsically related to its number of antenna panels and their associated baseband processing chains. In particular, the 3GPP specifies several multi-panel user equipment assumptions (MPUE-Assumption) depending on the number of baseband processing chains and the number of antenna panels at the MT side~\cite{3gpp.WI.NR_feMIMO}.\footnote{Note that, in this paper, we use the terms the \textit{multi-panel MT} and the \textit{directional MT} interchangeably in that a panel at the MT corresponds to a directional beam.} 

For instance, MPUE-Assumption \#1 covers the case when the MT has multiple antenna panels but only a single baseband processing chain to execute SSB measurements on them. This requires that the MT iterates over each single antenna panel---one at a time---during SSB bursts for conducting SSB measurements. This iterative process incurs additional delays for executing SSB measurements over all MT antenna panels. On the other hand, MPUE-Assumption \#3 covers the case where the MT has multiple antenna panels but each antenna panel has its own baseband processing chain. Thus the MT is capable of conducting SSB measurements on all antenna panels in a single SSB burst, hence resulting in a smaller measurement delay in comparison to MPUE-Assumption \#1. In short, the total average overhead per unit time depends on whether each of $2^k$ beams (interchangeably, antenna panels) at the typical MT has its own baseband processing chain or not. We discuss three possible cases as follows:

\subsubsection{Limited Overhead Case}
This case corresponds to MPUE-Assumption \#3, where each of the $2^k$ beams at the typical MT has its own baseband processing chain for independently conducting SSB measurements. Thus the SSB beam measurement can be done in all the beams of the MT simultaneously. In this case, the total average overhead per unit time is \begin{align}
T_{\mathrm{o}}(n) =  \mu_{\rm b}(n)T_{\mathrm{b}} +\mu_{\rm c} T_{\mathrm{c}}.
\label{eq:no_o}
\end{align}
Note that, in spite of $2^k$ beams at the typical MT, the average overhead per unit time is the same as if the MT is omnidirectional. But such a gain comes at the cost of increased hardware complexity at the MT due to the need of a baseband processing chain for each beam.

\subsubsection{Full Overhead Case}
This case corresponds to MPUE-Assumption \#1, where there is a single baseband processing unit that needs to be shared by all of the $2^k$ beams at the typical MT. Thus the SSB measurement for each of the $2^k$ beams is done sequentially. In this case, the total average overhead per unit time is 
\begin{align}
T_{\mathrm{o}}(n) =  2^k\left(\mu_{\rm b}(n)T_{\mathrm{b}} +\mu_{\rm c} T_{\mathrm{c}}\right).
\label{eq:full_o}
\end{align}
Here, the average overhead per unit time is $2^k$ times the one in the limited overhead case, but with an advantage of significantly less hardware complexity at the MT. 

A key disadvantage of the full overhead case is that the beam measurement is done during a BS handover even if the current serving BS and the next serving BS lie in the coverage of the same MT beam. Thus we propose a probabilistic beam switch case that uses the network geometry, namely, BS locations to reduce the overhead compared to the full overhead case.

\subsubsection{Probabilistic Beam Switch}
\begin{figure}
\centering
\includegraphics[scale=2]{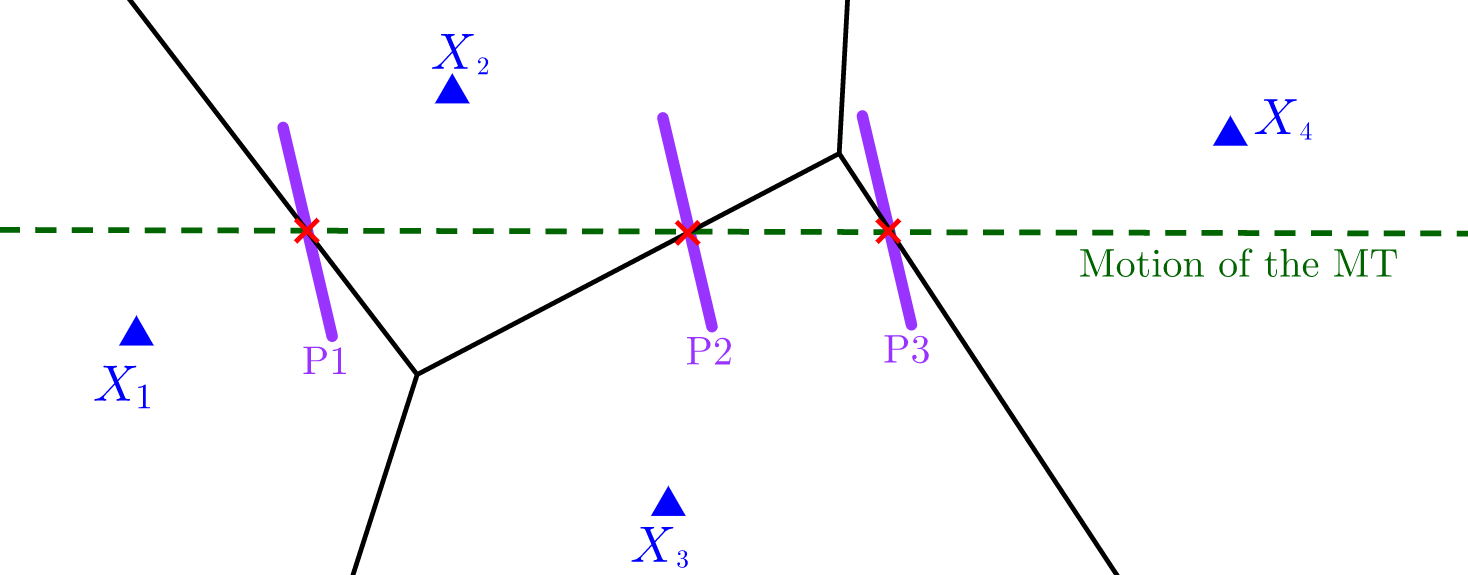}
\caption{Probabilistic beam switching with two beams (panels) at the MT. The triangles depict the BSs located according to a PPP. The solid black lines denote cell boundaries. A ``cross" denotes the BS handover location of the MT. $\rm{P1}$, $\rm{P2}$, and $\rm{P3}$ denote randomly oriented panels of the MT pertaining to each BS handover location, where each of the two beams covers the left and right halves separated by the panel. For instance, the BSs $X_1$ and $X_2$ lie in the coverage of two different beams, i.e., on the opposite side of the panel $\rm{P1}$. Thus there will be a beam switch during the handover from $X_1$ to $X_2$. On the contrary, $X_2$ and $X_3$ lie in the coverage of the same beam (the \textit{left} beam). Thus there is no need of a beam switch during the handover from $X_2$ to $X_3$.}
\label{fig:beam_switch}
\end{figure}

The idea is to predict the possibility of a beam switch during a BS handover since a beam switch is necessary if and only if the current and the next serving BSs lie in the coverage of different MT beams. The case of $2$ beams at the MT is depicted in Fig.~\ref{fig:beam_switch}. Exploiting the properties of the PPP that models BS locations, the probability that the current and the next serving BSs lie in different beams can be calculated~\cite{Delaunay_ang_note}. Specifically, when there are $2^k$ beams at the typical MT with main lobes oriented in uniformly at random directions and covering all directions, the probability of a beam switch is~\cite[Corollary 3]{Delaunay_ang_note}
\begin{align}
p_{\rm sw} = \frac{\pi}{2^k}\sin\frac{2^k}{\pi}, \qquad k \geq 1.
\label{eq:prob_sw}
\end{align}
Alternatively, only $p_{\rm sw}$ fraction of BS handovers require a beam switch.  When there are $2$, $4$, or $8$ beams at the MT, the probability of a beam switch during a BS handover is $0.637$, $0.9$, and $0.975$, respectively. As a result, the total average overhead per unit time can be given by
\begin{align}
T_{\mathrm{o}}(n) =  2^k\mu_{\rm b}(n)T_{\mathrm{b}} + (p_{\rm sw}2^k +1- p_{\rm sw})\mu_{\rm c} T_{\mathrm{c}}.
\label{eq:partial_o}
\end{align}

\section{Effective Area Spectral Efficiency}
\label{sec:effective_ASE}
For each of the three overhead cases discussed in Section~\ref{sec:cases_overhead}, the effective ASE per unit time is
\begin{align*}
\mathcal{R}_{\mathrm{eff}}(n, k) = \lambda (1-T_{\rm o}(n))^{+}\mathcal{R}(n,k),
\end{align*}
where $\mathcal{R}(n, k)$ is given by \eqref{eq:av_rate} and $(A)^{+} = \max(0, A)$.

For a given $k$, our objective is to find the integer $n$ that maximizes the effective ASE per unit time, i.e.,
\begin{align*}
n_{*} = \arg \max_{n \in \mathbb{N}} ~\mathcal{R}_{\mathrm{eff}}(n,k).
\end{align*}
The value of $n_{*}$ can easily be found by a linear search.

\section{Network Setup}
\label{sec:network_setup}
To validate our proposed model, we consider a $5$G NR-compliant radio access network (RAN) operating in a dense urban macro/pico cell scenario. A summary of the model parameters for FR1 (sub-$6 ~\mathrm{ GHz}$) and FR2 (above $6 ~\mathrm{GHz}$) network deployments is provided in Table~\ref{tab:1}. 

\begin{table}[!t]
\centering
\caption{Network parameter values for FR1 and FR2 deployments~\cite{3gpp.38.101-1,3gpp.38.101-2}}
\begin{tabular}{|l|l|l|}
\hline
  \textbf{Parameter}              &  \textbf{FR1}                    &  \textbf{FR2}                                            \\\hline
 Carrier frequency ($f_{\rm c}$)    &  $3.5~\mathrm{GHz}$                &  $28 ~\mathrm{GHz}                                        $ \\\hline
 Bandwidth ($W$)              &  $100 ~\mathrm{MHz}$ per carrier                &  $400~\mathrm{MHz}$ per carrier                                   \\\hline
 Noise density ($N_0$)           &  $-174 ~\mathrm{dBm/Hz}$            &  $-174 ~\mathrm{dBm/Hz}                                   $ \\\hline
 Transmit power ($P$)             &  $43~ \mathrm{dBm}$                 &  $36~\mathrm{dBm}                                        $ \\\hline
 Beam reselection overhead ($T_{\rm b}$)            &  $23~\mathrm{ms}$                   &  $23~\mathrm{ms}$                                           \\\hline
 Cell handover overhead ($T_{\rm c}$)           &  $43~\mathrm{ms} $                 &  $43~\mathrm{ms}$                                          \\\hline
 SSB periodicity ($\tau$)          &  $20~\mathrm{ms}$                  &  $20 ~\mathrm{ms                                        } $ \\\hline
 MT speed ($v$)            &  $[3, 30, 120]~\mathrm{km/h}$ &  $[3, 30]~\mathrm{km/h}                              $ \\\hline
 Inter-site distance (\rm{ISD})            &  $[250, 500, 1000] ~\mathrm{m}$  &  $[75, 125, 250{]} ~\mathrm{m}$                           \\\hline

 Maximum SINR ($Q_{\rm max}$)            &  $30 ~\mathrm{dB}$ &  $30 ~\mathrm{dB} $                          \\\hline
 Path loss exponent ($\alpha$)       &  3.5                    &  $\alpha_{\rm L} = 1.9$, $\alpha_{\rm N} = 3.5$  \\\hline
 Blockage model &  Implicit (NLOS)                   &  LOS ball                                      \\\hline
  LOS ball radius ($R_{\rm c}$)            &  $-$&  $75 ~\mathrm{m}                          $ \\\hline
\end{tabular}
\label{tab:1}
\end{table}

To explore the best potential of $5$G NR networks in operational bands FR1 and FR2, we use the maximum bandwidth allowed as per $5$G NR Release 15\cite{3gpp.38.101-1, 3gpp.38.101-2}, namely $100 ~\mathrm{MHz}$ per carrier for FR1 and $400~\mathrm{MHz}$ per carrier for FR2. 
For FR1, we assume NLOS conditions. Hence we set the value of path loss exponent $\alpha = \alpha_{\rm N}$ = 3.5. For FR2, when inside the LOS ball we assume $\alpha = \alpha_{\rm L}$ = 1.9, and we take $\alpha = \alpha_{\rm N}$ = 3.5 outside the LOS ball.

As the inter-site (or, equivalently inter-BS) distances (\rm{ISD}s) for FR1 are expected to be larger due to lower frequency-dependent path loss, we choose a transmission power $P$ = $43 ~\mathrm{dBm}$. For FR2, we can expect smaller \rm{ISD}s due to (a) higher attenuation loss at higher carrier frequencies, and (b) because massive MIMO and high beamforming gains imposes challenges in terms of RF exposure and EMF limitations. Thus we decrease $P$ to $36 ~\mathrm{dBm}$. We assume a noise level $N_0$ of -174 $\mathrm{
dBm/Hz}$, in order to represent a realistic scenario that is not limited by noise.

An important parameter is the intensity $\lambda$ of BSs, which is related to the average cell size and the \rm{ISD}. For an intensity $\lambda$ of BSs, the average cell size is $1/\lambda$. The average cell radius is defined as the radius $r_{\rm cell}$ of a ball having the same average area as the cell: $1/\lambda$. Then, the average \rm{ISD} is $2r_{\rm cell} = 2/(\sqrt{\pi \lambda})$. Hence we simply adjust the value of the intensity $\lambda$ of BSs to represent different \rm{ISD} scenarios. For FR1 bands, we analyze the following ISDs: $1000 ~\mathrm{m}$, $500 ~\mathrm{m}$, and $250~\mathrm{m}$. As for FR2 bands, the effect of propagation attenuation requires us to decrease the \rm{ISD}. Hence we consider the following \rm{ISD}s: $250 ~\mathrm{m}$, $125 ~\mathrm{m}$, and $75 ~\mathrm{m}$.

One important network planning decision is the number of beams per BS and per MT. In our model, this is related to the choice of the values for $n$ and $k$, respectively. The main and side lobe gains depend on the number $2^n$ and $2^k$ of beams. Specifically, an increase in the number of beams increases the main lobe gain and decreases the side lobe gain. Thus, without loss of generality, we assume the following antenna gains. The main lobe gain is $G_{{\rm m}, n} = 2^n$ at a BS ($G_{{\rm m}, k} = 2^k$ at the MT), while the side lobe gain is $G_{{\rm s}, n} = \frac{1}{2^n}$ at a BS ($G_{{\rm s}, k} = \frac{1}{2^k}$ at the MT).\footnote{As seen from previous sections, our model is amenable to different modeling of antenna gain patterns.}


\section{Numerical Results: Omnidirectional MT}
\label{sec:omni_result}

\begin{figure} 
    \centering
  \subfloat[\footnotesize \textbf{FR1}: Dashed line: $v = 3~\rm{km/h}$,
Dashed-dotted line: $v = 30~\rm{km/h}$, and
Solid line: $v = 120~\rm{km/h}$. \label{fig:diff_ISD_diff_vel_FR1}]{%
       \includegraphics[width=.7\linewidth]{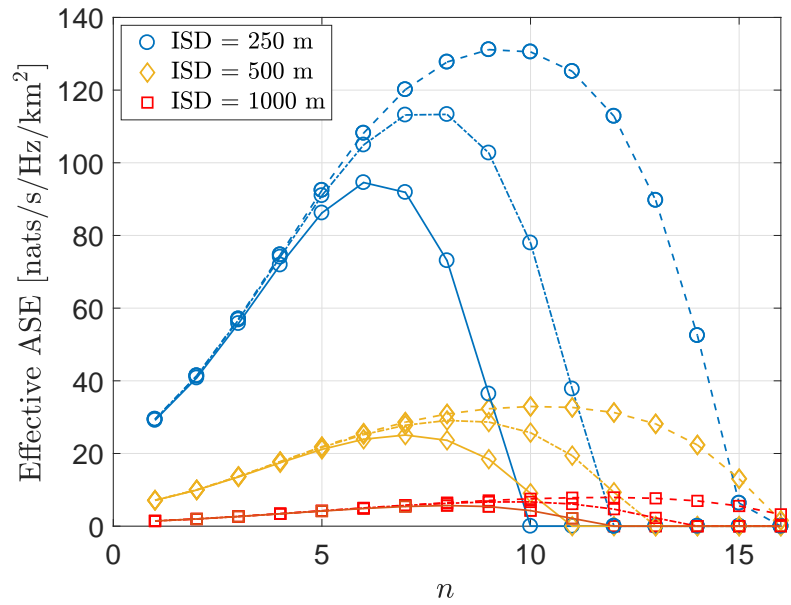}}
    \hfill
  \subfloat[\footnotesize \textbf{FR2}: Dashed line: $v = 3~\rm{km/h}$ and 
Solid line: $v = 30~ \rm{km/h}$.\label{fig:diff_ISD_diff_vel_FR2}]{%
        \includegraphics[width=0.7\linewidth]{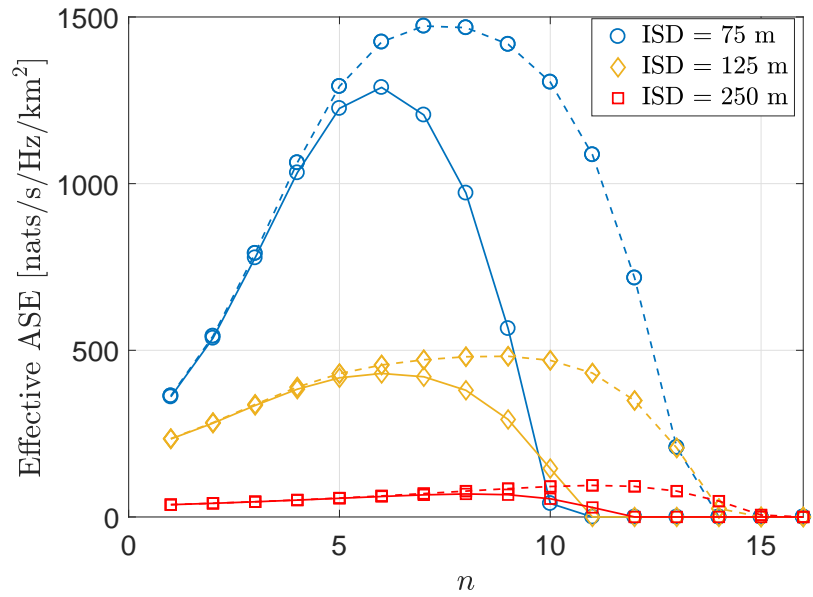}}
 
  \caption{Effect of average inter-site distance (\rm{ISD}) on the effective ASE for different $n$.}
  \label{fig:FR1_FR2_ISD} 
    
\end{figure}

We now demonstrate how the proposed model can be used for system-level evaluation of the effect of different system parameters on 5G NR network deployments over FR1 and FR2 for the \textit{omnidirectional} MT
for different MT speeds $v$ and \rm{ISD}s (in Figs.~\ref{fig:diff_ISD_diff_vel_FR1},  \ref{fig:diff_ISD_diff_vel_FR2}, and \ref{fig:comp_FR1_FR2}). In Section~\ref{sec:direct_result}, we will consider the case of the directional MT that has $2^k$, $k \geq 1$ beams. The discussions in this section pertain to Figs. \ref{fig:diff_ISD_diff_vel_FR1}, \ref{fig:diff_ISD_diff_vel_FR2}, and \ref{fig:comp_FR1_FR2}.

\subsection{Effect of the Number of Beams at BSs } 

In both FR1 and FR2 deployments, for a given \rm{ISD}, as $n$ (and hence the number $2^n$ of beams) increases, the beamforming gain increases and the interference from interfering BSs decreases due to narrower beams. But, at the same time, for a given $v$, the typical MT performs more frequent beam reselections due to a smaller beamwidth, increasing the time overhead. The negative impact of the increased time overhead gradually dominates the beamforming gain. Also, an increase in $n$ increases the probability of beam misalignment at the BS (given by \eqref{eq:p_bm}) as the MT is more likely to move into the beamwidth of another beam between two consecutive SSB measurements. As a result of this tussle between the increased beamforming gain, increased time overhead due to beam handovers, and the increased probability of beam misalignment, we observe that there is a value $n_{*}$ of $n$ that maximizes the effective ASE.

\subsection{Effect of the MT Speed}
With an increase in $v$, the MT crosses beam and cell boundaries more frequently resulting in a higher number of beam reselections and BS handovers, respectively. Also, the probability of beam misalignment increases. Hence, with an increase in $v$, the optimal value $n_{*}$ decreases, and for a fixed $n$, the effective ASE decreases.

\begin{figure}
\centering
\includegraphics[scale=.7]{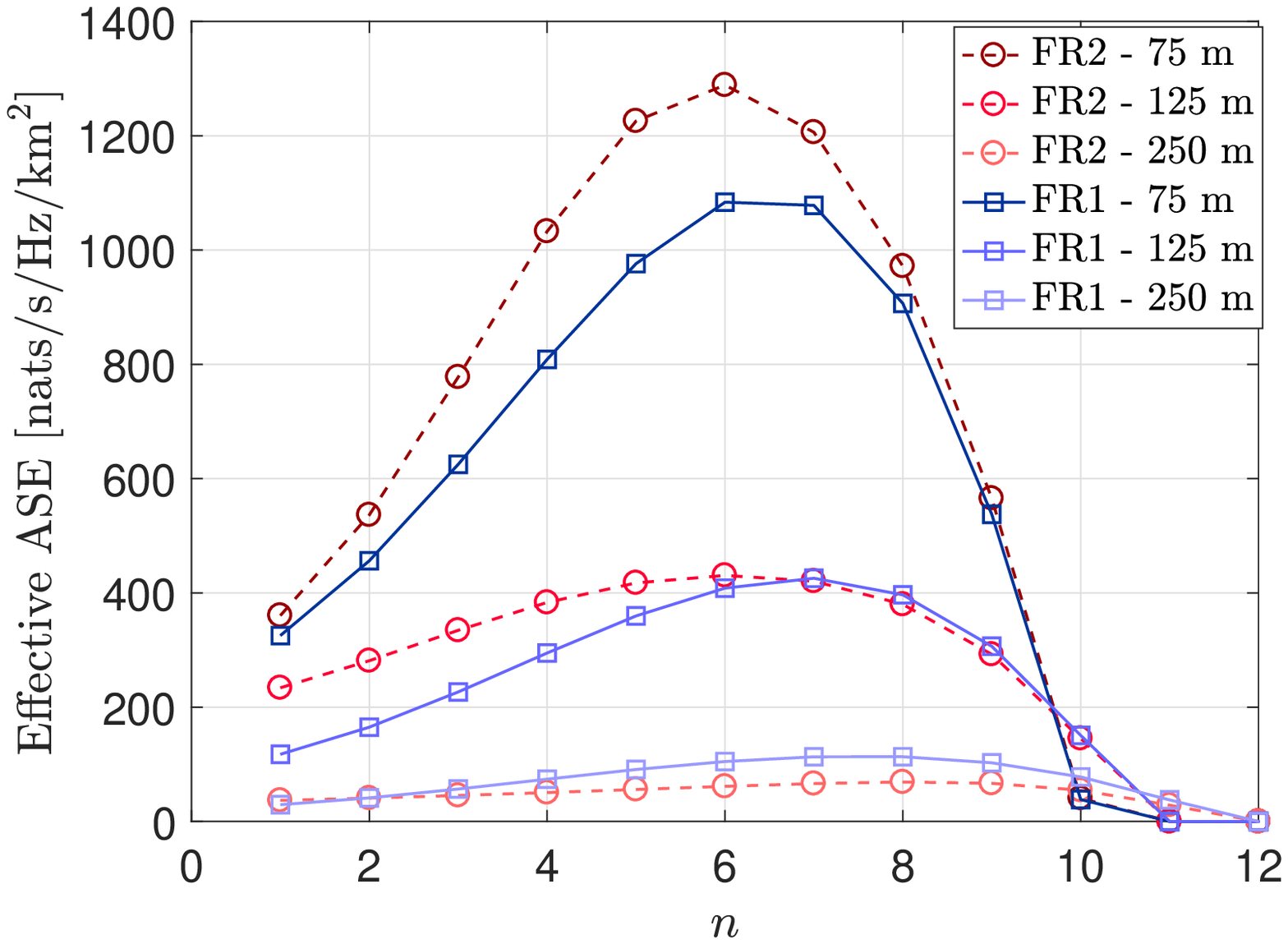}
\caption{Comparison between FR1 and FR2 deployments for different \rm{ISD}s with $v  = 30~\rm{km/h}$.}
\label{fig:comp_FR1_FR2}
\end{figure}

\subsection{Effect of the \rm{ISD}}
\label{sec:eff_ISD}
Note that a smaller \rm{ISD} means a denser cellular network. As the \rm{ISD} increases, the average cell size increases. Thus, BSs need to increase the number $2^n$ of beams for a higher beamforming gain (equivalently, a smaller beamwidth) and smaller interference. Hence, the value of $n_{*}$ increases with \rm{ISD}. For a smaller $n$, 
the time overhead associated with beam reselections is small. Hence, a smaller \rm{ISD} results in a higher effective ASE due to a smaller distance between the typical MT and its serving BS. This dominates the negative effects of increased interference power due to a denser deployment of BSs and increased time overhead due to more frequent BS handovers. But, for a large $n$, both beam reselections and BS handovers happen more frequently for a smaller \rm{ISD}, and the network with a larger \rm{ISD} achieves a higher effective ASE.

\subsection{Comparison of FR1 and FR2 Deployments}

When comparing beamwidths for FR1 with those for FR2 given the same $n$ value, we can expect that, even though the angular beamwidth for a given $n$ is the same, the linear width of the beams will be larger in FR1 than that in FR2 due to a \textit{usually} larger cell radius in the former.\footnote{Although nothing prohibits small \rm{ISD}s for FR1 (e.g., Wi-Fi operation in $2.4~\mathrm{ GHz}$ and $5~\mathrm{ GHz}$), pico/femto cell deployment models are the best fit for mmWave deployments.} Thus, it is expected that MTs operating in FR1 are less penalized due to overheads associated with beam reselections and BS handovers. Also, due to a larger \rm{ISD}, the \textit{distance-based} path loss when operating in FR1 is higher than when operating in FR2. Hence for the same value of $n$, we can expect that the interference in FR1 is smaller than that in FR2 despite a bit higher transmit power in FR1. This indicates that the additional received interference from a higher number of beams will not be as intense in FR2 as in FR1. As shown in Fig.~\ref{fig:comp_FR1_FR2}, this behavior is captured by our model, where the optimal value $n_{*}$ for the \rm{ISD} $= 250 ~\mathrm{m}$ in FR1 ($n_{*} = 8$) is higher than that in FR2 ($n_{*} = 6$) with \rm{ISD} $=75~\rm{m}$ or in FR2 with \rm{ISD} $=125~\rm{m}$ ($n_{*} = 7$).

Also, as Fig.~\ref{fig:comp_FR1_FR2} shows, when we compare the performances of FR1 and FR2 deployments for the same \rm{ISD}, we observe an interesting tradeoff. Smaller \rm{ISD}s (e.g., $75~\rm{m}$) benefit FR2 irrespective of the value of $n$ as, for a given critical LOS distance $R_{\rm c}$, it is more likely that the serving BS has LOS propagation to the MT (so smaller path loss) boosting signal power. Recall that, in FR1, even the serving BS always has NLOS propagation to the MT. As a result, the impact of increased time overheads with $n$ on the effective ASE is less in FR2 than that in FR1. As the \rm{ISD} increases, the probability of serving BS lying outside the LOS ball increases, in turn, reducing signal power significantly in FR2. Thus, when combined with higher path loss at higher frequencies in FR2, for higher values of $n$, the impact of time overheads is relatively higher in FR2 than FR1. Note that, as a result of the tussle between competing effects, the values of $n_{*}$ in FR1 and FR2 are close to each other for the same \rm{ISD}.

Overall, the numerical results  discussed in this section validate the claim that the proposed model allows for a system-level evaluation of network planning decisions in beam-based access networks such as $5$G NR. 

Note that, although the discussions on the effect of number of beams at the BS, MT speed, and the \rm{ISD} on the effective ASE have considered until now an omnidirectional MT, they are also valid for the case of directional MT. 

\section{Numerical Results: Directional MT}
\label{sec:direct_result}

We now study the effect of multiple beams at the typical MT and associated systemic tradeoffs. We will also compare the performance of the directional MT with that of the omnidirectional MT. Since it is expected that multiple beams at the MT will be used mostly in FR2 (mmWave) due to the need of directional communications, we restrict our discussion to FR2. 

\begin{figure}[!t]
    \centering
  \subfloat[\footnotesize Rayleigh fading\label{fig:rayl_FR2_ISD_75_vel_30}]{%
       \includegraphics[width=.7\linewidth]{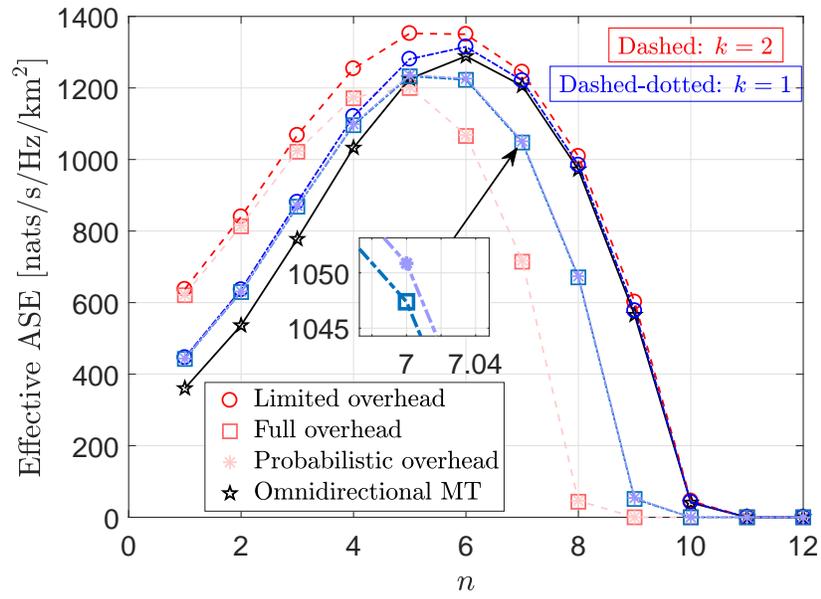}}
    \hfill
  \subfloat[\footnotesize Nakagami-$4$ fading\label{fig:naka_4_vel_30_ISD_75}]{%
        \includegraphics[width=0.7\linewidth]{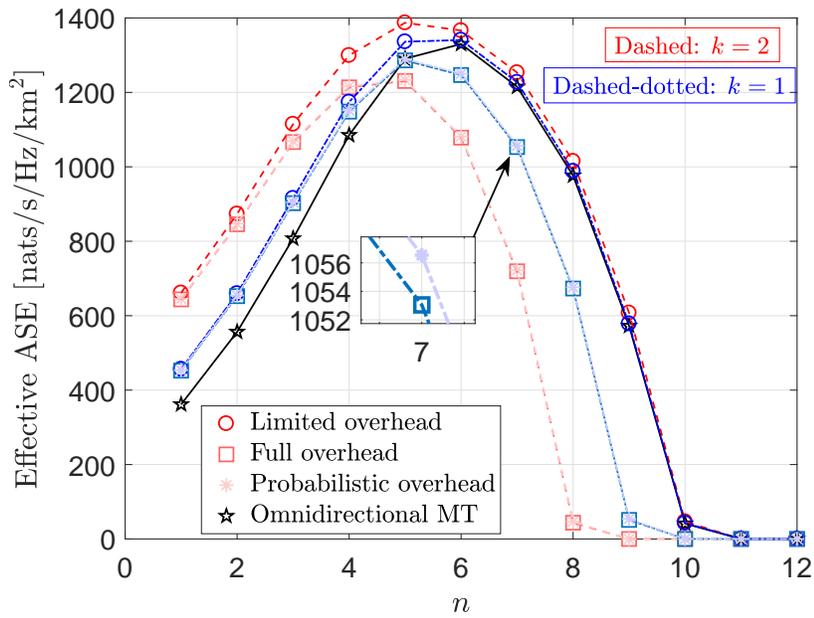}}
 
  \caption{Effect of the number $2^k$ beams at the typical MT on the effective ASE. $v  = 30~\rm{km/h}$ and $\rm ISD = 75$ m.}
  \label{fig:Ray_Naka_ISD_75_vel_30} 
    
\end{figure}

\begin{figure}
    \centering
  \subfloat[\footnotesize Rayleigh fading\label{fig:rayl_FR2_Rc_75_ISD_250_vel_30}]{%
       \includegraphics[width=.7\linewidth]{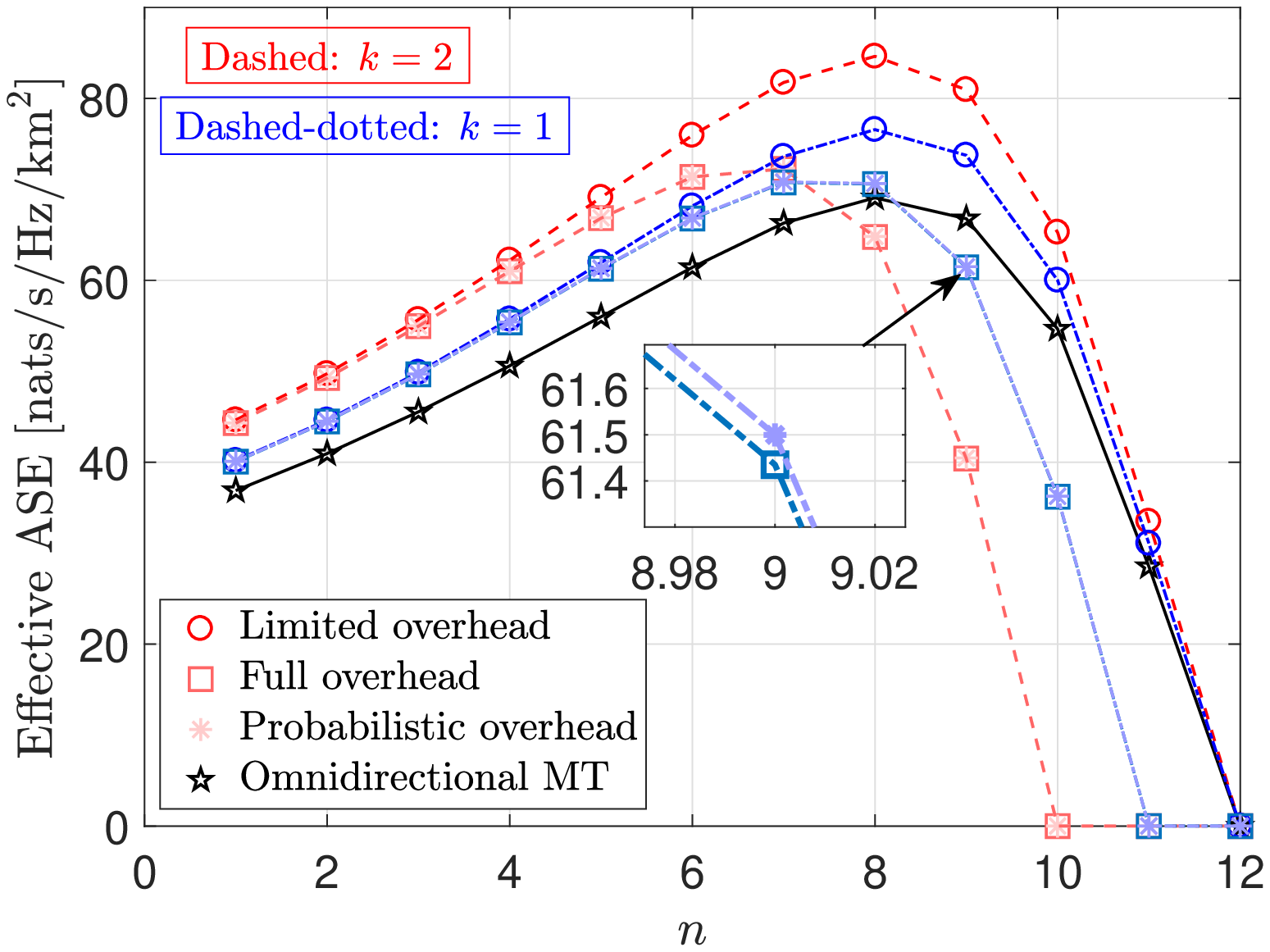}}
    \hfill
  \subfloat[\footnotesize Nakagami-$4$ fading\label{fig:naka_4_vel_30_ISD_250}]{%
        \includegraphics[width=0.7\linewidth]{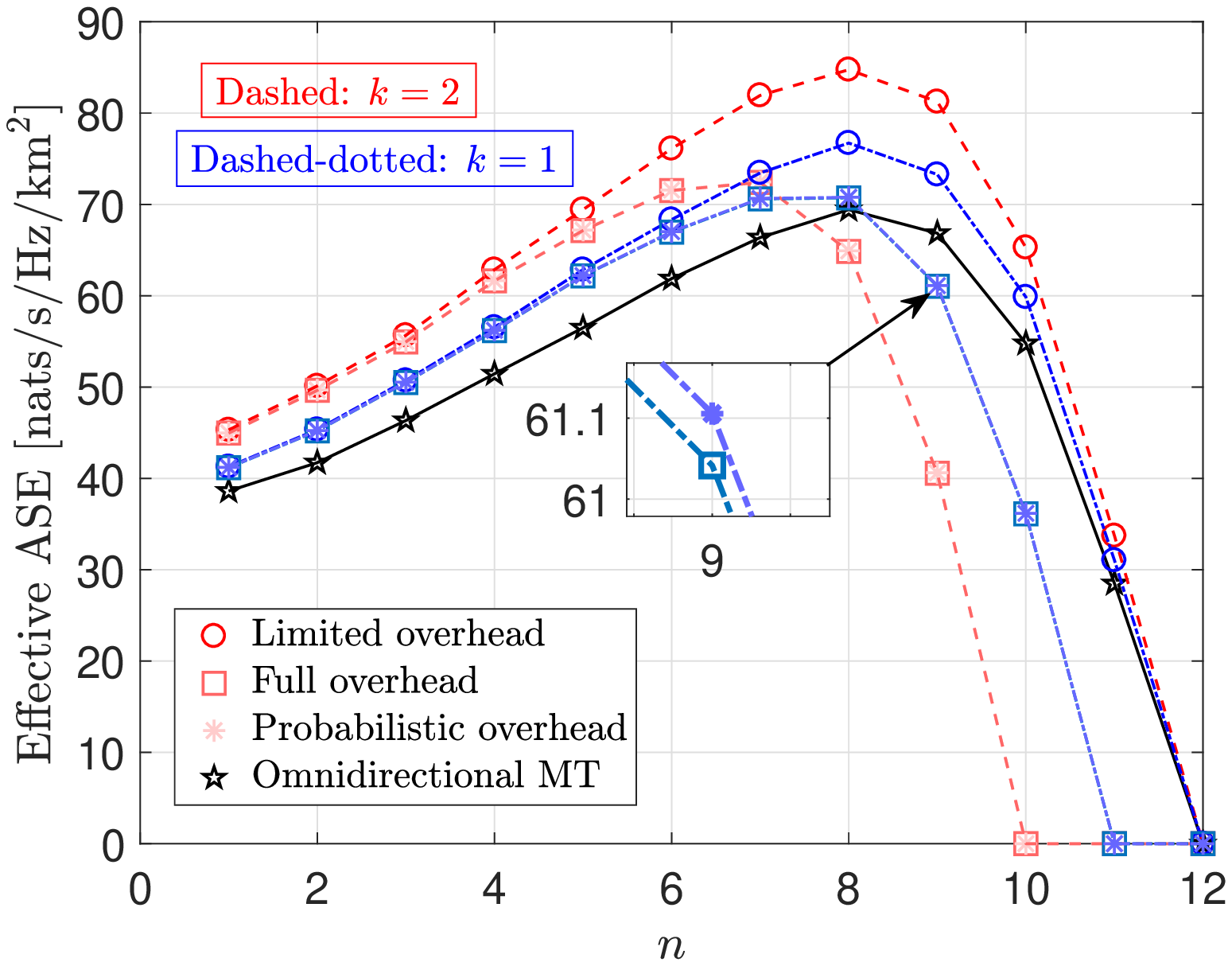}}
 
  \caption{Effect of the number $2^k$ beams at the typical MT on the effective ASE. $v  = 30~\rm{km/h}$ and $\rm ISD = 250$ m.}
  \label{fig:Ray_Naka_ISD_250_vel_30} 
  
\end{figure}

\subsection{Effect of the Number of Beams at the MT}

Figs.~\ref{fig:Ray_Naka_ISD_75_vel_30} and \ref{fig:Ray_Naka_ISD_250_vel_30} depict, for different \rm{ISD}s and fading models, the systemic tradeoffs associated with the directional MT and the effect of the number of baseband processing chains discussed in Section~\ref{sec:cases_overhead}. Recall that, as given by~\eqref{eq:no_o}, \eqref{eq:full_o}, and \eqref{eq:partial_o},  the time overheads for the \textit{limited overhead} (LO) case and the omnidirectional MT are the same, and are smaller when compared to the
\textit{probabilistic overhead} (PO) and \textit{full overhead} (FO) cases, with the FO case leading to the highest time overhead.
Specifically, Figs.~\ref{fig:Ray_Naka_ISD_75_vel_30} and \ref{fig:Ray_Naka_ISD_250_vel_30} compare the performance of the directional MT with $2$ ($k = 1$) and $4$ ($k = 2$) directional beams with the omnidirectional MT. For the LO case, the additional beamforming gain due to multiple directional beams at the MT (in addition to the beamforming gain at the BS) dominates consistently over the omnidirectional MT case, as the time overhead due to multiple beams at the MT remains constant for the LO case irrespective of the number $2^k$ of beams at the MT. In other words, for all values of $n$, the LO case with 4 beams at the MT performs best, and the LO case with 2 beams outperforms the omnidirectional MT case. Although the probability of beam misalignment at the MT increases with $k$ (see \eqref{eq:p_bm_MT}), the beamforming gain dominates this performance deteriorating factor.

On the other hand, for the FO case,  the time overhead increases proportionally to both the number $2^k$ of beams at the MT as well as the number $2^n$ of beams at the BS (see \eqref{eq:full_o}). Thus, after a certain value of $n$, the beamforming gain at the MT is eventually dominated by the increased time overhead and the probability of beam misalignment at both the MT and the BS. For instance, as shown in Fig.~\ref{fig:rayl_FR2_ISD_75_vel_30}, for $2^5 = 32$ beams at the BS, the case with $4$ beams at the MT achieves better effective ASE compared to the case with $2$ beams at the MT, and the trend reverses from $2^6 = 64$ beams onwards at the BS. Also, for the FO case, when the number $2^n$ of beams at the BS is large enough (e.g., 32 beams in Fig.~\ref{fig:rayl_FR2_ISD_75_vel_30}), the advantage of the directional communication at the MT vanishes due to the increased time overhead, and the omnidirectional reception at the MT leads to the better effective ASE. The optimal value of $n$ that maximizes the effective ASE reduces with an increase in $k$ in order to balance between the positive effect of the beamforming and the negative impact of the increased time overhead and the increased probability of beam misalignment. For a given number of beams at the MT, i.e, for fixed $k$, the PO case offers an advantage over the FO case due to a smaller number of beam switching at the MT for the given set of network parameters.

To determine the effect of fading on the performance, we compare two fading models: Rayleigh and Nakagami-$m$. For the latter, we assume that the LOS propagation is characterized by Nakagami-$4$ fading. As shown in Figs.~\ref{fig:Ray_Naka_ISD_75_vel_30} and \ref{fig:Ray_Naka_ISD_250_vel_30}, both fading models present the same performance trend in the effective ASE for different number of beams at the BS and the MT, \rm{ISD}s, and the overhead cases. This confirms the interest of our mathematical results based on Rayleigh fading.

As Figs.~\ref{fig:Ray_Naka_ISD_75_vel_30} and \ref{fig:Ray_Naka_ISD_250_vel_30} together show, the effect of the \rm{ISD} for the directional MT is the same as that for the omnidirectional MT (discussed in Section~\ref{sec:eff_ISD}). With respect to the directional MT case, the comparison of Figs.~\ref{fig:Ray_Naka_ISD_75_vel_30} and \ref{fig:Ray_Naka_ISD_250_vel_30} reveals that the crossover point, where the FO case with 4 beams at the MT ($k = 2$) starts performing worse compared to the FO case with 2 beams at the MT ($k = 1$) and to the omnidirectional MT, shifts towards right with an increase in \rm{ISD} as the influence of the beamforming gain increases due to larger average cell sizes.

\begin{figure} 
    \centering
  \subfloat[\footnotesize Limited overhead case\label{fig:no_three_ISD_200_vel_30}]{%
       \includegraphics[width=.7\linewidth]{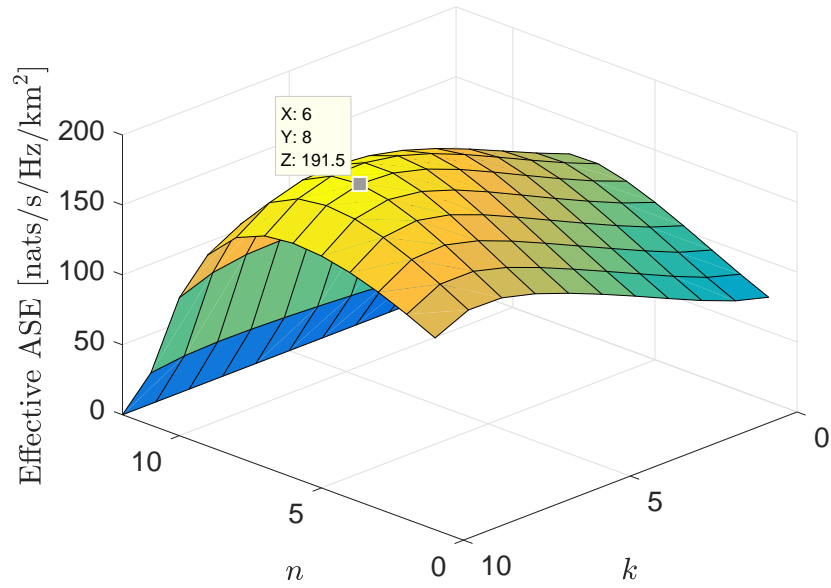}}
    \hfill
  \subfloat[\footnotesize Full overhead case\label{fig:full_three_vel_30_ISD_200}]{%
        \includegraphics[width=0.7\linewidth]{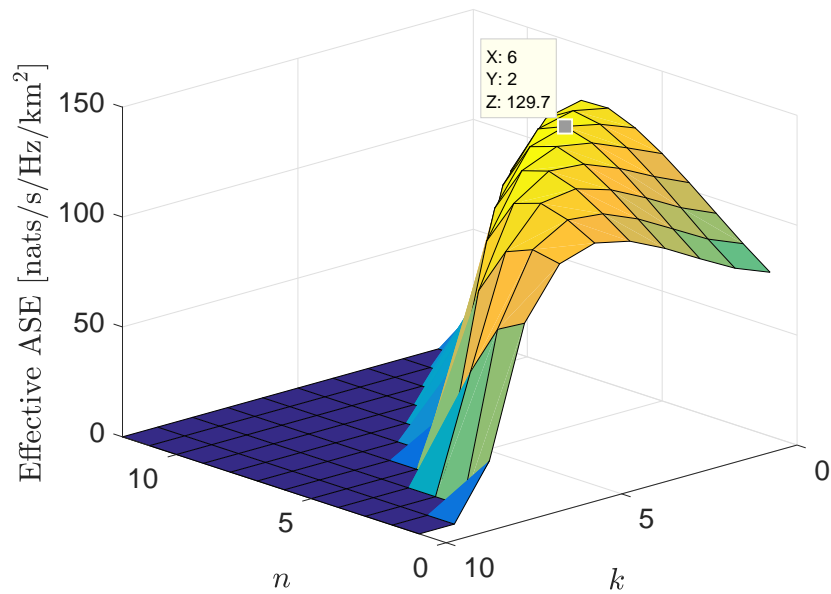}}
  \caption{Three-dimensional plot of the effective ASE against the number $2^n$ beams at the BS and the number $2^k$ beams at the MT. $v  = 30~\rm{km/h}$ and $\rm ISD = 200$ m.}
  \label{fig:three_ISD_200_vel_30} 
    
\end{figure}

Fig.~\ref{fig:three_ISD_200_vel_30} gives a general view of the effect of directional communications from the perspectives of both the BS and the MT. Specifically, it gives the optimal number of beams at the BS as well as the MT using a 2-D numerical optimization. For the LO case, the optimal number of beams at the MT is $2^8 = 256$, while it is $2^6 = 64$ at the BS. This is due to the fact that, for the LO case, the time overhead does not increase with the number of beams at the MT. Thus, it is better to increase the number of beams at the MT instead of at the BS to achieve sufficient beamforming gain. However, the negative impact of increased probability of beam misalignment eventually dominates, which reduces the effective ASE after increasing the number of beams beyond $256$ at the MT.\footnote{As of now, having $256$ beams at the MT may be infeasible, but it may be a reality in the future, especially for MTs that are not smartphones, such as flying drones and vehicles.} On the other hand, as Fig.~\ref{fig:full_three_vel_30_ISD_200} shows, the optimal number of beams at the MT reduces to $4$ for the FO case as a result of the proportional increase in the time overhead with the number of beams at the MT. 

\begin{figure}

	\centering
			\includegraphics[scale=0.68]{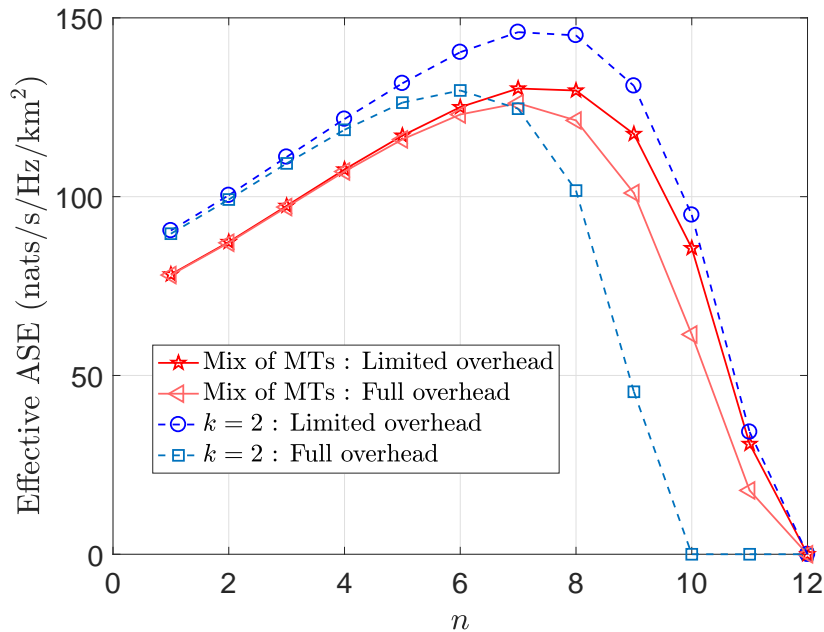}
			\caption{Heterogeneous vs. homogeneous networks. $v  = 30~\rm{km/h}$ and $\rm ISD =200$ m. $p_0 = 0.6$, $p_1 = 0.3$, and $p_2 = 0.1$.}
			\label{fig:comp_het_homo_Rayl}
		
\end{figure}

\subsection{Network Performance for the Mix of MTs}
\label{sec:mix_MT}

Fig.~\ref{fig:comp_het_homo_Rayl} compares a homogeneous network consisting of only MTs with $4$ beams ($k = 2$) and a heterogeneous network consisting of a mix of MTs traveling at the same speed of $v = 30~\rm{km/h}$. For the heterogeneous network, $60\%$ of the MTs are omnidirectional ($p_0 = 0.6$), $30\%$ of the MTs are with $2$ beams ($p_1 = 0.3$), and $10\%$ of the MTs are with $4$ beams ($p_2 = 0.1$). The performance for the mix of the MTs can be obtained by weighing the ergodic Shannon rate given by~\eqref{eq:av_rate} by the probabilities $p_0$, $p_1$, and $p_2$.

Specifically, for the LO case, the homogeneous network of MTs with $4$ beams suffers from a loss in the effective ASE if other types of MTs are to be added in that network. This is an expected result as the MTs with $2$ beams and the omnidirectional MTs achieve a smaller beamforming gain, while the time overhead due to multiple beams at the MT remains the same irrespective of the type of the MT. On the other hand, for the FO case, the design of the network is more complicated. In particular, Fig.~\ref{fig:comp_het_homo_Rayl} depicts how the number of beams at the BS must be adjusted depending on the composition of the network. For instance, when the network consists of only MTs with $4$ beams, having a small number beams at the BS is beneficial from the effective ASE point of view. On the contrary, if the network is heterogeneous as per the aforementioned composition, having a large number of beams at the BS is beneficial.

\begin{figure}
	\centering
			\includegraphics[scale = 0.68]{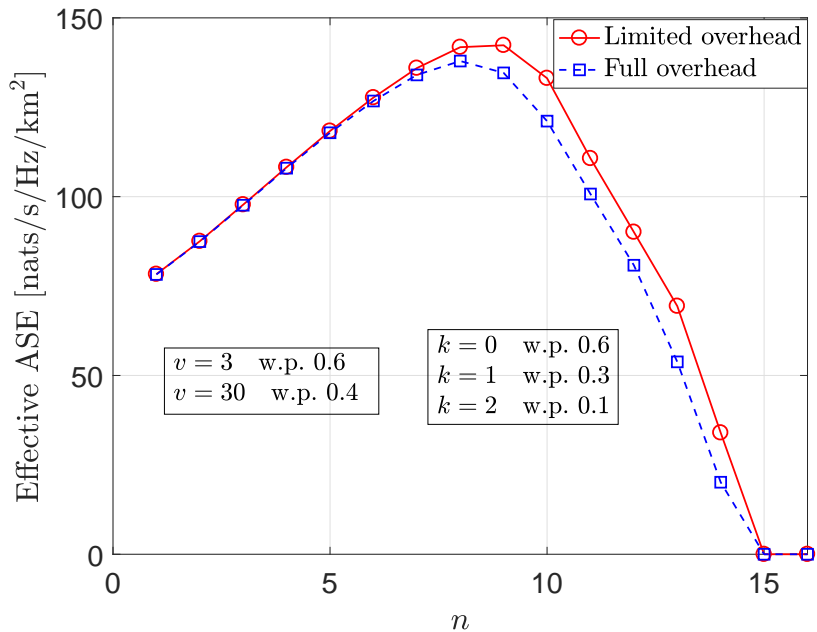}
			\caption{Overall performance for the mix of MTs. $\rm ISD =200$ m. $p_0 = 0.6$, $p_1 = 0.3$, and $p_2 = 0.1$. $q_3 = 0.6$ and $q_{30} = 0.4$.}
			\label{fig:overall_ASE_vel_panels_p_0_6_0_3_0_1_vel_3_0_6_30_0_4_ISD_200}
	
\end{figure}

Fig.~\ref{fig:overall_ASE_vel_panels_p_0_6_0_3_0_1_vel_3_0_6_30_0_4_ISD_200} captures the overall performance of the network consisting of a mix of MTs where $60\%$ of MTs move at $3~{\rm km/h}$ ($q_3 = 0.6$), while $40\%$ move at $30~{\rm km/h}$ ($q_{30}  = 0.4$). For each type of MT in terms of speed, $60\%$ of the MTs are omnidirectional ($p_0 = 0.6$), $30\%$ of the MTs have $2$ beams ($p_1 = 0.3$), and $10\%$ of the MTs are with $4$ beams ($p_2 = 0.1$). In such a network, $2^9 =512$ beams are required at the BS to maximize the effective ASE for the LO case, while $2^8 = 256$ beams are sufficient for the FO case. Note that, despite of large time overhead in the FO case compared to the LO case, the optimal effective ASE achieved in the former case ($137.9$ nats/s/Hz/$\rm{km}^2$) is close to that in the latter case ($142.3$ nats/s/Hz/$\rm{km}^2$). Recalling that the FO case has reduced hardware complexity compared to the LO case, Fig.~\ref{fig:overall_ASE_vel_panels_p_0_6_0_3_0_1_vel_3_0_6_30_0_4_ISD_200} implies that the network operator might be able to provide almost the same quality of service to MTs at reduced hardware complexity for certain network compositions.

\section{Study of Some New Performance Metrics}
\label{sec:new_metrics}
In this section, we investigate how new performance metrics obtained in this paper and given in Table~\ref{tab:2} behave. These metrics are of independent interest.

\subsection{Probability of Beam Misalignment at the BS}

\begin{figure} 
    \centering
  \subfloat[\footnotesize \rm{ISD} = 25, 75, 250, 500, 1000~m (left to right), $v$ = 30~km/h, $\tau$ = 20~ms.\label{fig:p_bm_BS_ISD_v_30_tau_20}]{%
       \includegraphics[width=.32\linewidth]{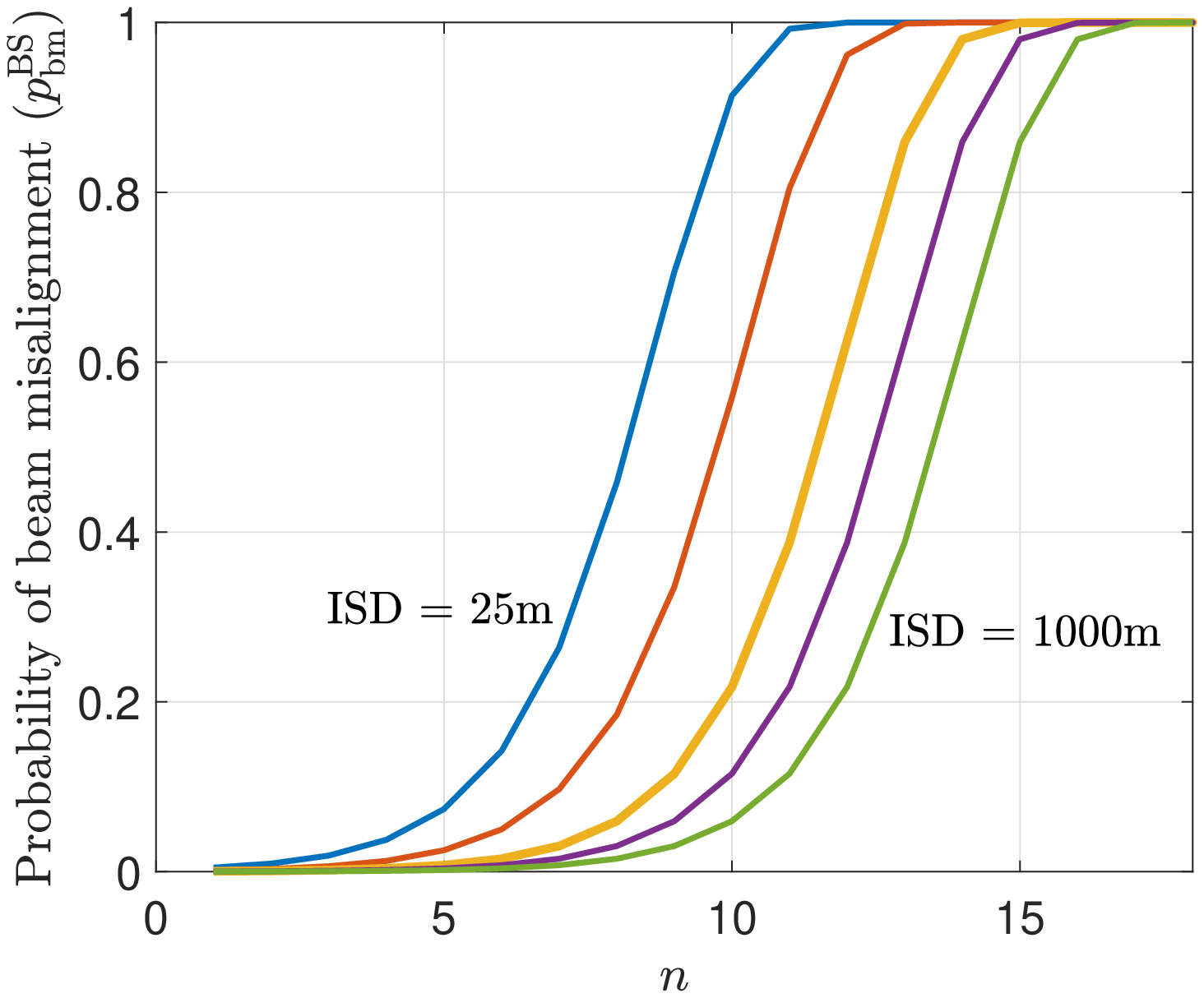}}
    \hfill
  \subfloat[\footnotesize $v$ = 3, 30, 120, 250~km/h (right to left), \rm{ISD} = 200~m, $\tau$ = 20~ms.\label{fig:p_bm_v_ISD_200_tau_20}]{%
        \includegraphics[width=0.32\linewidth]{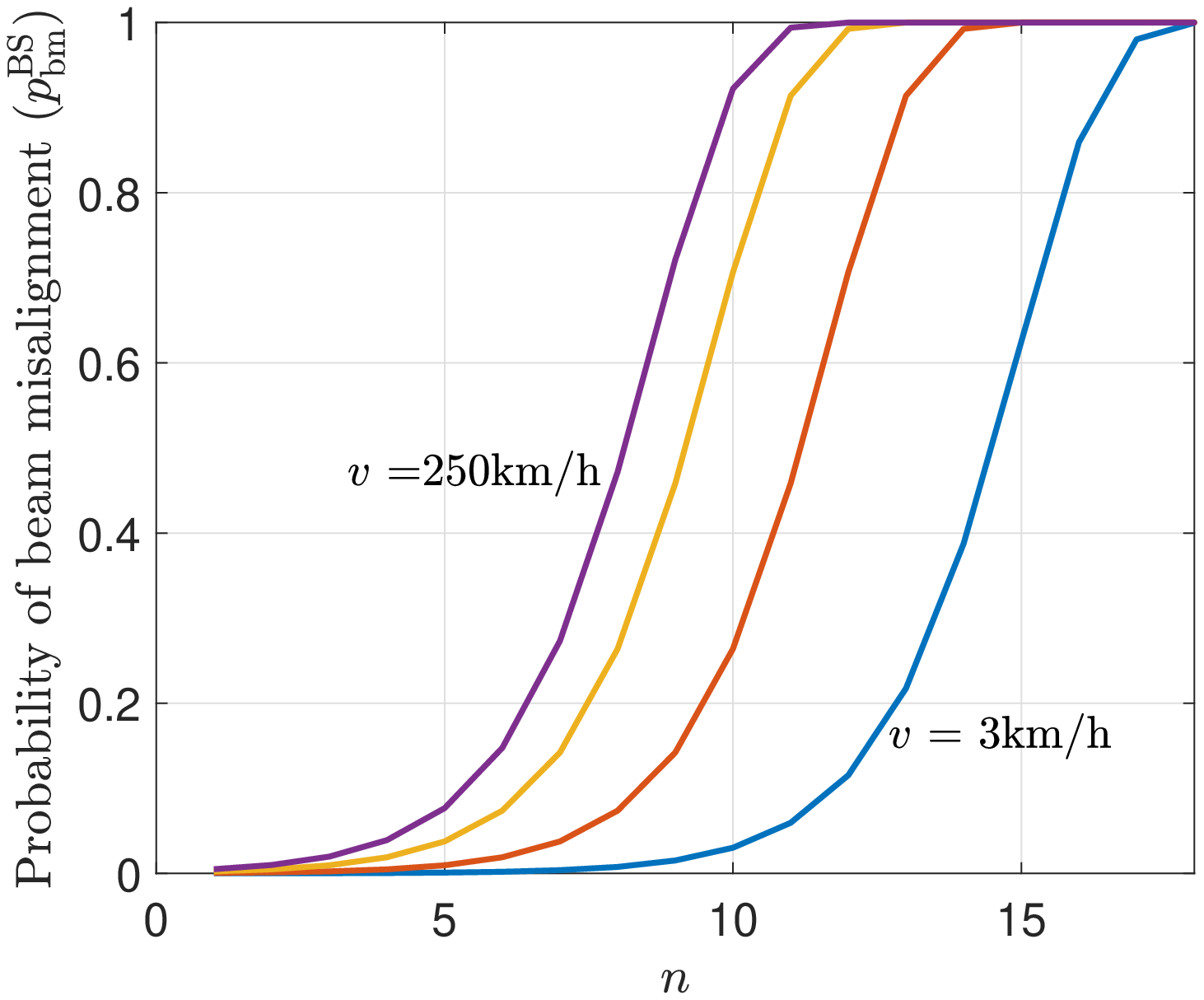}}
     \hfill
  \subfloat[\footnotesize $\tau$ = 5, 10, 20, 40, 80, 160~ms (right to left), \rm{ISD} = 200~m, $v$ = 30~km/h. \label{fig:p_bm_v_tau_ISD_200_v_30}]{%
        \includegraphics[width=0.32\linewidth]{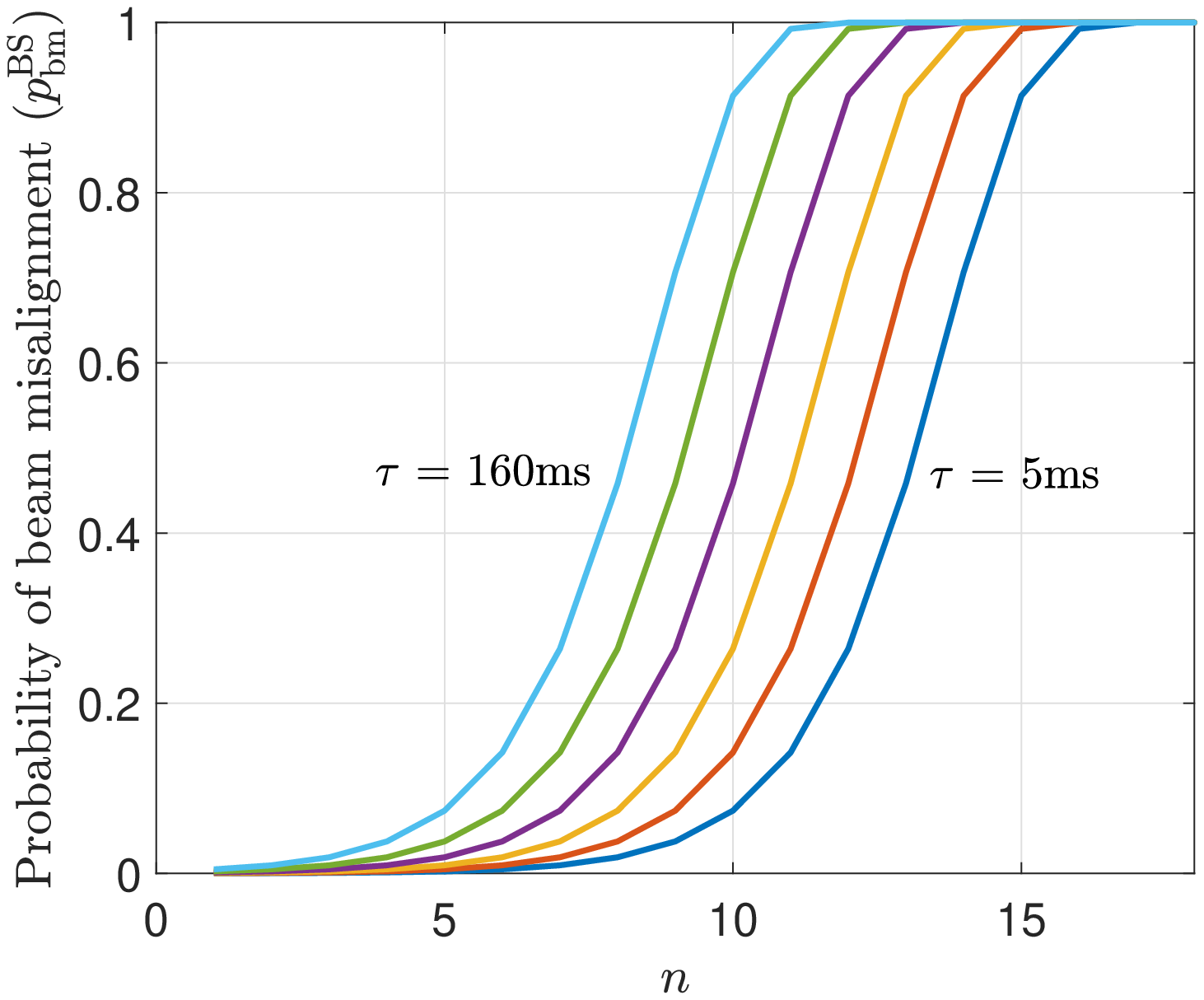}}
  \caption{Effect of the ISD, MT speed, and the SSB periodicity on the probability of beam misalignment at the BS.}
  \label{fig:p_bm} 
    
\end{figure}

Fig.~\ref{fig:p_bm} depicts how the probability of beam misalignment $p_{\rm bm}^{\rm BS}$ at the BS behaves with the ISD, MT speed $v$, and the SSB periodicity $\tau$ as a function of $2^n$ beams at the BS. As the network becomes denser with a decrease in the ISD, $p_{\rm bm}^{\rm BS}$ increases since the average cell size decreases, which in turn increases the chances that the MT moves in the coverage of the new beam between two consecutive SSBs without performing a beam reselection (Fig.~\ref{fig:p_bm_BS_ISD_v_30_tau_20}). Thus, for a dense 5G NR network, such as with \rm{ISD} = $25~\rm{m}$, having a large number beams at the BS might not offer much advantage as the beamforming gain is quickly dominated by the negative impact of the beam misalignment as $n$ increases. Also, the faster the MT moves, the higher are the chances that the MT moves in the coverage of the new beam between two consecutive SSBs without performing a beam reselection (Fig.~\ref{fig:p_bm_v_ISD_200_tau_20}). Finally, Fig.~\ref{fig:p_bm_v_tau_ISD_200_v_30} shows that increasing the SSB periodicity results in a longer duration between two consecutive SSB bursts, which hence increases $p_{\rm bm}^{\rm BS}$ at the BS. A similar analysis could be performed to analyze the effect of the ISD, MT speed, and the SSB periodicity on the probability of beam misalignment at the MT.

\begin{figure}
		\begin{center}
			\includegraphics[scale = 0.68]{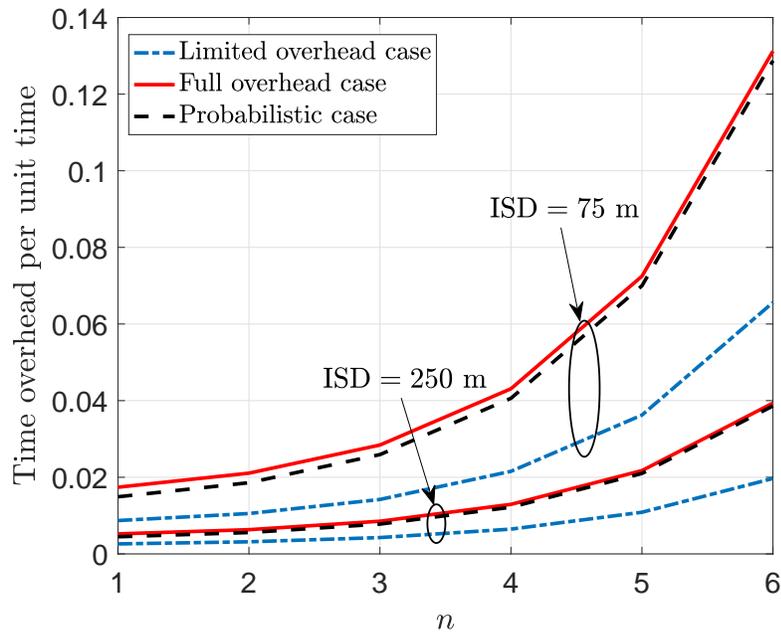}
			\caption{Time overhead per unit time vs. the number of beams at the BS. $v = 30~\rm{km/h}$, $k = 1$ (i.e., 2 beams at the MT), $\tau = 20~\rm{ms}$.}
			\label{fig:to_MT_panels}
		\end{center}
\end{figure}

\subsection{Time Overheads}

As discussed in Section~\ref{sec:cases_overhead}, the time overheads due to beam selections during BS handovers and beam reselections due to the intra-cell mobility of the MT crucially depend on the number of baseband processing units at the MT. Fig.~\ref{fig:to_MT_panels} depicts the effect of the number of beams at the BS on the time overhead. For smaller ISD values, the average cell size is smaller, which leads to more frequent BS handovers and beam reselections within the cell. This in turn increases the time overhead per unit time. Due to the same reason, the difference between the full overhead case and the probabilistic overhead case is larger at a smaller ISD. Thus the latter case is expected to bring more gains in terms of the time for data transmissions in ultradense cellular networks.

When the number of beams at the BS is smaller (say $2^3 = 8$), the difference between the three overhead cases is smaller. Thus they all lead to almost the same effective ASE as seen in Figs.~\ref{fig:Ray_Naka_ISD_75_vel_30} and \ref{fig:Ray_Naka_ISD_250_vel_30}. But the cases of full and probabilistic overhead have lesser hardware complexity due to a single baseband processing unit compared to two units in the limited overhead case. On the other hand, for a large number of beams at the BS, the time overhead in the limited overhead case is much smaller, which may justify having a higher hardware complexity.

\section{Conclusions and Future Directions}
\label{sec:conclusions}
This paper has presented a first system-level mathematical framework for beam management in $5$G RAN. This framework allows one to capture essential scenario characteristics (e.g., operational frequency, blockage characteristics), technological features (e.g., beamforming configuration, delay overheads), network deployment choices (e.g., \rm{ISD}, carrier bandwidth), and traffic scenarios (e.g., mix of MTs). Also, it leads to the definition of several new key performance metrics proper to 5G NR RANs, and stochastic geometry was used to derive closed-form expressions of each of them. These expressions are then aggregated in a global formula allowing one to evaluate the effective area spectral efficiency of this 5G RAN. This effective area spectral efficiency is the Shannon rate per Hertz averaged out over all users of a large network, when taking into account both the pros of beam densification in terms of improved SINR, and its cons in terms of the handover overheads that are incurred by mobile users when they switch beams or cells. For a dense urban macro/pico cell scenario in sub-6 GHz and mmWave bands, the framework accurately captures the systemic tradeoffs between \rm{ISD}s, path loss, interference, and signaling overhead due to beam management. This also shows that the stochastic geometry model can be effective for conducting system-level analysis of beam-based radio access networks such as $5$G NR.

In the future, the model can be improved to incorporate sectorized BSs with antenna panels and their effects on beam shaping/gains, MT blockage models, and others. Further, our model can be extended to capture other tradeoffs involved in designing a beam-based RAN. For instance, the average beam time-of-stay, i.e., the time a user remains with a given beam before switching to another beam, decreases with an increase in the number of beams in the beamset, which impacts the available time for performing channel estimation, link adaptation, and power control. Also, decreasing the periodicity interval in which beamformed reference signal resources are monitored may improve the effectiveness and response time of beam refinement, but at the same time it results in reduced spectral efficiency due to the control overhead.

\appendices

\section{Proof of Lemma~\ref{lem:suc_prob}}
\label{app:suc_prob}

The success probability is
\begin{align}
p_{\rm s}(n, k, \beta) & \triangleq \mathbb{P}(\mathsf{SINR}_{n,k} > \beta).
\label{eq:ps_app}
\end{align}
Based on whether the serving BS lies within the LOS ball of radius $R_{\rm c}$ or not, we can write the success probability $p_{\rm s}$ as
\begin{align}
p_{\rm s}(n,k,\beta) &= \int_{0}^{R_{\rm c}} \mathbb{P}(\mathsf{SINR}_{n,k} > \beta \mid r)f_{R}(r)\mathrm{d}r + \int_{R_{\rm c}}^{\infty} \mathbb{P}(\mathsf{SINR}_{n, k} > \beta \mid r)f_{R}(r)\mathrm{d}r,
\end{align}
where $f_{R}(r) = 2\pi \lambda r \exp(-\lambda \pi r^2)$ is the probability density function of the distance $R$ of the typical MT to the nearest BS. When $0 < r < R_{\rm c}$, we have
\begin{align}
\mathbb{P}(\mathsf{SINR}_{n,k} > \beta \mid r) &= \mathbb{P}\left(
\frac{PKG_0 h_{X_0}r^{-\alpha_{\rm L}}}{N_0 + I_{n,k}} > \beta \,\middle\vert\,  r\right) \nonumber \\
&= \mathbb{P}\left(h_{X_0} > \frac{\beta r^{\alpha_{\rm L}} (N_0 + I_{n,k})}{PKG_0} \,\middle\vert\,  r\right) \nonumber \\
& = e^{-\frac{\beta r^{\alpha_{\rm L}}N_0}{PKG_0}}\mathbb{E}\left[\exp\left(-\frac{\beta r^{\alpha_{\rm L}} I_{n,k}}{PKG_0}\right)\right]
\end{align}
for a given $G_0$ which is a random variable with the probability mass function given by \eqref{eq:gain_serv}.

We have
\begin{align}
\mathbb{E}\left[\exp\left(-\frac{\beta r^{\alpha_{\rm L}} I_{n,k}}{PKG_0}\right)\right] & = \mathbb{E}\left[\exp\left(-\frac{\beta r^{\alpha_{\rm L}} \sum_{X \in \Phi\setminus \lbrace X_0\rbrace} P G_X h_{X}l(|X|)}{PKG_0}\right)\right] \nonumber \\
&\overset{(\mathrm{a})}{=} \mathbb{E}\left[\prod_{X \in \Phi\setminus \lbrace X_0\rbrace} \mathbb{E}\left(\exp\left(-\frac{\beta r^{\alpha_{\rm L}} G_X h_{X}l(|X|)}{KG_0}\right)\right)\right] \nonumber \\
&\overset{(\mathrm{b})}{=}\mathbb{E}\left[\prod_{X \in \Phi\setminus \lbrace X_0\rbrace} \mathbb{E}\left(\frac{1}{1 + \frac{\beta r^{\alpha_{\rm L}}l(|X|)G_X}{KG_0}}\right)\right]\nonumber \\
& \overset{(\mathrm{c})}{=} \mathbb{E}\left[\prod_{X \in \Phi\setminus \lbrace X_0\rbrace} \left(\frac{p_{\rm mm}}{1 + \frac{\beta r^{\alpha_{\rm L}}l(|X|)G_{{\rm m}, n}G_{{\rm m}, k}}{KG_0}} + \frac{p_{\rm ms}}{1 + \frac{\beta r^{\alpha_{\rm L}}l(|X|)G_{{\rm m}, n}G_{{\rm s}, k}}{KG_0}} \right. \right. \nonumber \\
&\left. \left. + \frac{p_{\rm sm}}{1 + \frac{\beta r^{\alpha_{\rm L}}l(|X|)G_{{\rm s}, n}G_{{\rm m}, k}}{KG_0}} + \frac{p_{\rm ss}}{1 + \frac{\beta r^{\alpha_{\rm L}}l(|X|)G_{{\rm s}, n}G_{{\rm s}, k}}{KG_0}}\right)\right], 
\end{align}
where $(\mathrm{a})$ follows from the i.i.d. nature of fading random variables, $(\mathrm{b})$ follows from averaging over the channel power gains on interfering channels, and $(\mathrm{c})$ follows from the probability mass function of the gain $G_X$ (given by \eqref{eq:gain_intf}) of an interfering BS at $X \in \Phi$.

From the probability generating functional (PGFL) of PPP, it follows that
\begin{align}
\mathbb{E}\left[\exp\left(-\frac{\beta r^{\alpha_{\rm L}} I_{n,k}}{PKG_0}\right)\right] &= \exp\left(-2\pi \lambda \int_{r}^{\infty} \left(1- \frac{p_{\rm mm}}{1+\frac{\beta r^{\alpha_{\rm L}} l(w)G_{{\rm m}, n} G_{{\rm m}, k} }{KG_0 }} -  \frac{p_{\rm ms}}{1+\frac{\beta r^{\alpha_{\rm L}} l(w)G_{{\rm m}, n} G_{{\rm s}, k} }{K G_0}} \right.\right. \nonumber \\ 
&\left.\left. -  \frac{p_{\rm sm}}{1+\frac{\beta r^{\alpha_{\rm L}} l(w)G_{{\rm s}, n} G_{{\rm m}, k} }{KG_0}} -  \frac{p_{\rm ss}}{1+\frac{\beta r^{\alpha_{\rm L}} l(w)G_{{\rm s}, n} G_{{\rm s}, k} }{K G_0}} \right)w \mathrm{d}w\right) \nonumber \\
&= \exp\!\left(\!-\int_{r}^{R_{\rm c}}F(\alpha_{\rm L}, \alpha_{\rm L}, w, G_0) \mathrm{d}w - \int_{R_{\rm c}}^{\infty}\!F(\alpha_{\rm L}, \alpha_{\rm N}, w, G_0)\mathrm{d}w\!\right),
\label{eq:main_side}
\end{align}
where, in \eqref{eq:main_side}, the first integral corresponds to interfering BSs within the LOS ball and the second integral to those outside the LOS ball with  
$F(\alpha_{\rm S}, \alpha_{\rm I}, w, G_0)$ given by \eqref{eq:F}.

Similarly, one can obtain an expression for $\mathbb{P}(\mathsf{SINR}_n > \beta \mid r)$ when $R_{\rm c} \leq r < \infty$, where the interfering BSs lie outside the LOS ball. By averaging over the distance $R$ of the typical MT to its nearest BS, one gets \eqref{eq:suc_prob}. Finally, averaging over the gain $G_0$ from the serving BS yields the final expression of the success probability in \eqref{eq:suc_prob_main}.

\section{Proof of Theorem~\ref{thm:int_beam}}
\label{app:int_beam}
\begin{figure}
\centering
\includegraphics[scale=.43]{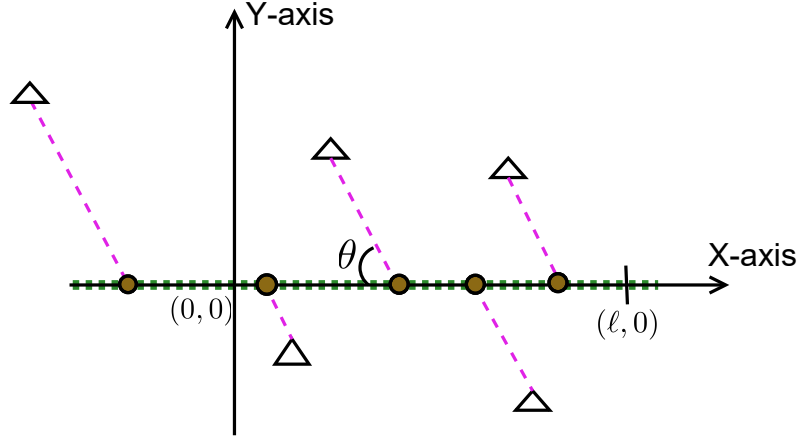}
\caption{At most one geometry-based beam reselection for a BS. $\bigtriangleup$ : Base station locations, brown circle : Beam reselection location, dashed lines : Beam boundaries, and dotted line : MT trajectory along the $x$-axis.}
\label{fig:BM_model}
\end{figure}

The geometry-based beam reselections occur at beam boundaries of a BS. In Fig.~\ref{fig:PVC}, these locations of beam reselections are denoted by ``brown circles". Let $\theta$ denote the angle of a beam boundary with respect to the direction of the motion of the MT (see Fig.~\ref{fig:BM_model}). This angle $\theta$ is distributed uniformly at random in $[0,\pi]$. In Fig.~\ref{fig:BM_model}, a `brown circle' denotes a point along the $x$-axis where the typical MT reselects the beam. Let $\Psi \subset \mathbb{R}$ denote the point process of beam reselection events, where the points of $\Psi$ are indicated by `brown circles'. We are interested in calculating the average number of beam reselections the typical MT performs per unit length, which is equivalent to calculating the intensity of $\Psi$.

Without loss of generality, consider the motion of the typical MT in the interval $[0,\ell]$, which corresponds to the motion of the typical MT from $(0, 0)$ to $(\ell, 0)$ along the $x$-axis. First, we calculate the average number of beam reselections when there is at most one beam reselection corresponding to a BS. This happens when there are two beams of the same size for a BS. Using the result for two beams, we can obtain the intensity of beam reselections corresponding $2^n$ beams at a BS.

\begin{figure}
\centering
\includegraphics[scale=.37]{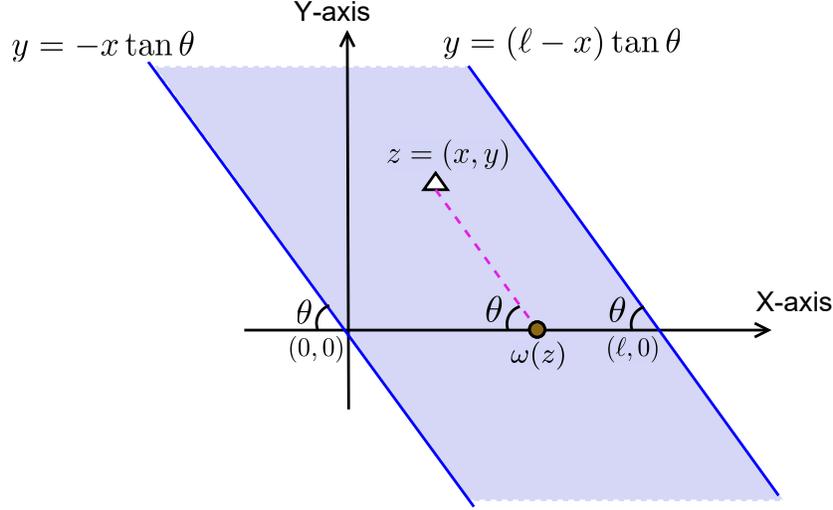}
\caption{Strip $\mathcal{S}$ (shaded area) between two blue solid lines with angle $\theta$ with the $x$-axis. Triangle : Base station location, brown circle : Beam reselection location, and dashed line : Beam boundary.}
\label{fig:strip}
\end{figure}

During the interval $[0,\ell]$, the typical MT may move through the Voronoi cells of multiple BSs. As shown in Fig.~\ref{fig:strip}, let $\omega(X)$ denote the point of the beam reselection corresponding to the BS located at $X \in \Phi$. The event of beam reselection corresponding to a BS at $X\in \Phi$ occurs when the following two events occur simultaneously:
\begin{enumerate}

\item the point of the beam reselection lies in the Voronoi cell of the BS at $X\in \Phi$, i.e., $\omega(X) \in V_{\Phi}(X)$ where $V_{\Phi}(X)$ is the Voronoi cell of the BS located at $X \in \Phi$.
\item the point of the beam reselection lies on the line connecting $(0, 0)$ and $(\ell,0)$, i.e, $\omega(X) \in [0,\ell]$. 
\end{enumerate}
Consequently, conditioning on $\theta$, the average of number of beam reselections in $[0,\ell]$ is
\begin{align*}
\mathbb{E}(\Psi[0,\ell]\mid \theta)  = \mathbb{E}\left(\displaystyle\sum_{X \in \Phi} \boldsymbol{1}_{\omega(X) \in V_{\Phi}(X)}\boldsymbol{1}_{\omega(X) \in [0,\ell]}\right),
\end{align*}
where $\boldsymbol{1}$ is the indicator function. Conditioning on the fact that there is a BS at location $z \in \mathbb{R}^2$ and using the Campbell's theorem~\cite{FB_book}, it follows that
\begin{align}
\mathbb{E}(\Psi[0, \ell]\mid \theta)  =\lambda \int_{\mathbb{R}^2}\mathbb{P}_{z}(\omega(z) \in V_{\Phi}(z))~\boldsymbol{1}_{\omega(z) \in [0, \ell]}~\mathrm{d}z, 
\label{eq:campbell}
\end{align}
where $\mathbb{P}_z(\cdot)$ denotes the Palm probability.

\textbf{Event $\boldsymbol{\omega(z) \in V_{\Phi}(z)}$}: This event occurs when there is no BS closer to the typical MT than the one at location $z$. This is equivalent to the event that there is no BS in the ball of radius $\|z - \omega(z)\|$ centered at $\omega(z)$. Since the BS point process is a PPP with intensity $\lambda$, we have
\begin{align}
\mathbb{P}_{z}(\omega(z) \in V_{\Phi}(z)) &= \exp\left(-\lambda \pi \|z - \omega(z)\|^2\right) \nonumber \\
&= \exp\left(-\frac{\lambda \pi y^2}{\sin^2 \theta}\right),
\label{eq:voronoi_event}
\end{align}
where $\|z - \omega(z)\| = \frac{y}{\sin \theta}$ since $z  = (x, y)$. 

\textbf{Event $\boldsymbol{\omega(z) \in [0,\ell]}$}: This event occurs if the BS at $z$ lies within the strip $\mathcal{S}$ between two lines passing through the origin $(0,0)$ and $(\ell,0)$ at angle $\theta$ as shown in Fig.~\ref{fig:strip}. Equivalently, by representing $z$ in Euclidean coordinates as $z = (x,y)$, it follows that 
\begin{align}
\boldsymbol{1}_{\omega(z) \in [0,\ell]}~\mathrm{d}z =\boldsymbol{1}_{z \in \mathcal{S}} ~\mathrm{d}z  =  \boldsymbol{1}_{(x,y) \in \mathcal{S}}~\mathrm{d}x\mathrm{d}y.
\label{eq:line_event}
\end{align}
The left line passing through the origin at an angle $\theta$ can be given as $y = -x\tan\theta$, while the right line passing through $(\ell,0)$ at an angle $\theta$ can be given as $y = (\ell-x)\tan \theta$. Thus, the BS at $z$ lies in the strip $\mathcal{S}$ if 
\begin{align}
-\infty < y < \infty ~~~\text{and}~~~ -\frac{y}{\tan \theta} \leq x \leq \ell-\frac{y}{\tan \theta}.
\label{eq:boundary_strip}
\end{align}

From \eqref{eq:voronoi_event}, \eqref{eq:line_event}, and \eqref{eq:boundary_strip}, we can express \eqref{eq:campbell} as
\begin{align*}
\mathbb{E}(\Psi[0,\ell]\mid \theta)  &= \lambda \int_{y = -\infty}^{\infty}\mathrm{d}y\int_{x = -\frac{y}{\tan \theta}}^{\ell -\frac{y}{\tan \theta}}\exp\left(-\frac{\lambda \pi y^2}{\sin^2 \theta}\right)\mathrm{d}x  \nonumber \\
&=\ell \left(\sqrt{\lambda} |\sin \theta|\right) .
\end{align*}

Since $\theta$ is uniformly distributed in $[0,\pi]$, averaging over $\theta$ yields
$
\mathbb{E}(\Psi[0,\ell])  = \left(\frac{2\sqrt{\lambda}}{\pi}\right)\ell.
$
Hence, for the case of two beams, the linear intensity of beam reselections is $\frac{2\sqrt{\lambda}}{\pi}$.

For $2^n$ beams with $n \in \mathbb{N}$, there are $2^{n-1}$ possibilities of beam reselections corresponding to the serving BS. In this case, we can obtain the average number of beam reselections by superimposing the beam reselection events corresponding to each beam boundary crossing considered earlier in this proof. Hence, the linear intensity of beam reselections for $2^n$ beams at a BS is
\begin{align}
\mu_{\rm s,b}(n) = 2^{n-1}\frac{2\sqrt{\lambda}}{\pi} =\frac{2^n\sqrt{\lambda}}{\pi}.
\label{eq:int_beam}
\end{align}
Considering the speed $v$ of the MT, one gets the time intensity $\mu_{\rm t,b}$ of geometry-based beam reselections as 
$
\mu_{\rm t,b}(n) = \frac{2^n\sqrt{\lambda}}{\pi}v.
$

\end{document}